%% file: TCI_arxiv.tex
\documentclass[journal,twoside,web]{IEEEtran}
\usepackage{cite}
\usepackage{amsfonts}
\usepackage{graphicx}
\usepackage{textcomp}
\usepackage[mathscr]{euscript}
\usepackage{algorithm,algpseudocode} 
\usepackage{times}
\usepackage{epsfig}
\usepackage{amsmath}
\usepackage{amssymb}
\usepackage{longtable}
\usepackage{lscape}
\usepackage{booktabs}
\usepackage[export]{adjustbox}%
\usepackage{comment}
\usepackage{pict2e}
\usepackage{gensymb}
\usepackage{graphics}
\usepackage{xspace}
\usepackage{textcomp}
\usepackage{bm}
\usepackage{diagbox}
\usepackage[mathscr]{euscript} 
\usepackage{multirow}
\usepackage{epstopdf}
\usepackage{multicol}
\usepackage{cite}
\usepackage{hyperref}
\usepackage{mathtools}
\usepackage{subfig}
\usepackage{color}
\newcommand{\SJ}[1]{\textcolor{black}{#1}}
\newcommand{\SL}[1]{\textcolor{black}{#1}}
\newcommand{\SD}[1]{\textcolor{black}{#1}}
\newcommand{\SR}[1]{\textcolor{black}{#1}}
\input{macro}

\begin{document}
\title{Adaptive Local Neighborhood-based Neural Networks for MR Image Reconstruction from Undersampled Data }

\author{Shijun Liang, \IEEEmembership{Student Member, IEEE},
Anish Lahiri, \IEEEmembership{Member, IEEE},  Saiprasad~Ravishankar,~\IEEEmembership{Senior~Member,~IEEE}
\thanks{© 2024 IEEE. Personal use of this material is permitted. Permission from IEEE must be obtained for all other uses, in any current or future media, including reprinting/republishing this material for advertising or promotional purposes, creating new collective works, for resale or redistribution to servers or lists, or reuse of any copyrighted component of this work in other works.}
\thanks{S. Liang is with the Department of Biomedical Engineering, Michigan State University, East Lansing, MI, 48824 USA (liangs16@msu.edu).}
\thanks{The work was done when A. Lahiri was with the Department of Electrical and Computer Engineering, University of Michigan, Ann Arbor, MI, 48901 USA
(anishl@umich.edu).}
\thanks{S. Ravishankar is with the Department of Computational Mathematics, Science and Engineering and the Department of Biomedical Engineering, Michigan State University, East Lansing, MI, 48824 USA (ravisha3@msu.edu).}
\thanks{This work was supported in part by a research gift from the Advanced Radiology Services (ARS) Foundation.
}}

\maketitle

\begin{abstract}
Recent medical image reconstruction techniques focus on generating high-quality medical images suitable for clinical use at the lowest possible cost and with the fewest possible adverse effects on patients. Recent works have shown significant promise for reconstructing MR images from sparsely sampled k-space data using deep learning. In this work, we propose a technique that rapidly estimates deep neural networks directly at reconstruction time by fitting them on small adaptively estimated neighborhoods of a training set. In brief, our algorithm alternates between searching for neighbors in a data set that are similar to the test reconstruction, and training a local network on these neighbors followed by updating the test reconstruction. Because our reconstruction model is learned on a dataset that is in some sense  
similar to the image being reconstructed rather than being fit on a large, diverse training set, it is more adaptive to new scans. It can also handle changes in training sets and flexible scan settings, while being relatively fast. Our approach, dubbed LONDN-MRI,
was validated on multiple data sets using deep unrolled reconstruction networks. Reconstructions were performed at four fold and eight fold undersampling of k-space with 1D variable-density random phase-encode undersampling masks. Our results demonstrate that our proposed locally-trained method produces higher-quality reconstructions compared to models trained globally on larger datasets as well as other scan-adaptive methods.

\end{abstract}


\section{Introduction}
\label{sec:introduction}
In applications like X-ray computed tomography (CT)~\cite{ct} and 
magnetic resonance imaging (MRI)~\cite{5484183}, reconstructing
images from undersampled or corrupted observations is of critical importance. 
For example, this is necessary to reduce a patient's exposure 
to radiation in CT or reduce time spent acquiring MRI data. 
MRI scans involve sequential data acquisition resulting in
long acquisition times that are not only a burden for patients and hospitals, but also make MRI susceptible to motion artifacts. 
Reconstructing images from limited measurements can speed up the MRI scan, but usually entails solving an ill-posed inverse problem. 
Recent approaches to accelerating MRI acquisition such as compressed 
sensing (CS)~\cite{compress} reduce scan time by collecting fewer measurements while preserving image quality by exploiting image priors or regularizers. 
Historically, regularization in CS-MRI has been based on 
sparsity of wavelet coefficients~\cite{wave} or using total variation~\cite{totalv}. While conventional CS assumes sparsity of signals in known bases, approaches based on learned image models have been shown to be more effective for MRI reconstruction, starting with learned synthesis dictionaries~\cite{ravishankar2011dlmri,jacob2013blindCSMRI}. \SJ{The dictionary parameters could be learned from unpaired clean image patches from a dataset and used for reconstruction or learned simultaneously with image reconstruction~\cite{SOUP-DIL,CTdict,super,ravishankar2011dlmri}.} Additionally, recent advances in sparsifying transform learning have resulted in efficient or inexpensive data-adaptive sparsity-based reconstruction frameworks for MRI~\cite{ravishankar2012learning,ravishankar2020review,wensailukebres19}. Other contemporary techniques could allow learning explicit regularizers in a supervised manner~\cite{9747201} for improved image restoration.

\SR{Deep learning (DL) has emerged as a potent methodology for tackling large-scale inverse problems, notably in enhancing image reconstruction techniques in MRI and CT. Predominantly, end-to-end CNN, as exemplified by the U-net model~\cite{Unet,jin:17:dcn}, have been employed to mitigate artifacts arising from undersampling in MRI data. Additionally, a plethora of alternative network models such as the Transformer~\cite{transfomer2021task}, and Generative Adversarial Networks (GANs)~\cite{Gans}, have demonstrated their effectiveness in MRI reconstruction, as detailed in comprehensive reviews like~\cite{ravishankar2020review}. Furthermore, transfer learning~\cite{transfer_learning_MRI} 
has also been used with neural networks for MRI reconstruction to achieve domain transfer.}

\SR{To enhance both stability and performance, hybrid-domain approaches such as~\cite{modl} enforce data consistency (i.e., the reconstruction is enforced to be consistent with the measurement model) all through 
 training and reconstruction. Networks incorporating data consistency layers are pivotal in MR imaging for maintaining alignment between the reconstructed image and the original data in k-space~\cite{Zheng2019twodataconsist,casade2017deep}. This category encompasses various methodologies, including deep unrolling-based methods~\cite{sun2016deep,hammernik2018learning}(which adapt traditional iterative algorithms to learn regularization parameters), regularization by denoising approaches~\cite{romanoRED17}, and plug-and-play methods~\cite{buzzard:18:pap}, among others. Distinctively, the ADMM-CSNet~\cite{sun2016deep} utilizes neural networks for the optimization of ADMM parameters, diverging from the  ISTA-Net~\cite{zhang2018istanet}, which focuses on refining CS reconstruction models grounded in the Iterative Shrinkage-Thresholding Algorithm.  The 1D deep low-rank and sparse network (ODLS)~\cite{One_dimen_MRI} demonstrates enhanced robustness for 2D MR image reconstruction, particularly in scenarios characterized by a limited number of training samples. While these deep learning-based reconstruction methods have demonstrated superiority over traditional CS techniques, concerns regarding their stability and interpretability persist, as highlighted in~\cite{AIconcern}.}

Apart from algorithmic advances, another
driving force behind deep 
learning-based reconstruction is the 
the rapid growth of publicly available 
training datasets. The availability 
of (paired or unpaired) training data sets made possible by
efforts like OCMR~\cite{chen2020ocmr} and
fastMRI~\cite{zbontar2019fastmri} has 
enabled rapidly demonstrating the capacity of deep 
learning-based algorithms for improved image reconstruction or denoising in MRI applications.

However, one major drawback of these learned approaches is that they typically require large training datasets such as fully sampled MRI data to be effective. A recent scan-specific deep learning method is the deep image prior~\cite{ulyanov2018deep}, which has been applied to MRI~\cite{DIP_MRI} and learns a neural network for reconstruction in an unsupervised fashion from a single image's measurements. Other scan-adaptive methods include RAKI~\cite{Raki}, which is a nonlinear deep learning-based auto-regressive auto-calibrated reconstruction method. RAKI could be viewed as a deep neural network-based version of the parallel imaging scheme GRAPPA~\cite{Grappa}.  
LORAKI~\cite{loraki,lobos2018robust} is another scheme that trains an autocalibrated recurrent neural network (RNN) to recover missing k-space data.
All these methods learn scan-specific networks without requiring large datasets. A related approach dubbed self-supervised learning has also shown promise for MRI~\cite{akcakaya20} and uses a large unpaired data set. 

\subsection{Contributions}
While deep learning approaches have gained popularity for MRI reconstruction due to their ability to model complex data sets, they often have difficulties generalizing to new data or distinct experimental situations at test time.

Deep CNNs usually require enormous datasets for offline training to ensure adequate performance trade-offs.
In this work, we propose to learn adaptive LOcal NeighborhooD-based Networks for MRI (LONDN-MRI) reconstruction. The approach efficiently learns reconstruction networks from small clusters in a training set, directly at reconstruction time.

\begin{itemize}
    \item The proposed models are trained using a small number of adaptively chosen neighbors that are in proximity (or are similar in a sense) to the underlying (to be reconstructed) image (cf.~\cite{lahiri:20:csa} for a slightly related approach in the context of patch-based dictionary learning).
    
    \item We show connections of this algorithm to a challenging bilevel optimization problem. Our algorithm for image reconstruction alternates between finding a small set of similar images to a current reconstruction, training the network locally on such neighbors, and updating the reconstruction. 

    \item \SR{The proposed local learning approach is flexible and can be seamlessly integrated with various existing deep learning frameworks for MRI, such as unrolled networks and image-domain denoisers,  
    to enhance their performance.}

    \item \SR{Our experimental results on multiple datasets (fastMRI, Stanford FSE, and fastMRI+) and across multiple k-space undersampling factors show that the proposed local adaptation techniques
    surpass networks trained globally on larger datasets. We demonstrate improved performance against scan-specific deep learning methods such as deep image prior, RAKI, and LORAKI, even when using a small number of neighbors for training.}

    \item \SR{We have shown the method's generalizability under different scenarios including different sampling patterns, and testing on data with artificial as well as natural lesions, when the training dataset didn't include such lesions. 
    To establish clinical utility, we also conducted tests under different MR scan contrast settings and varying signal-to-noise ratios at test time, where the proposed method showed promise. Our study also encompassed an analysis of image quality vs. time consumption trade-offs when involving different networks and number of neighbors selected, and compared favorably with related approaches.}
\end{itemize}

\subsection{Organization}
The rest of this article is organized as follows. Section~\ref{section2} discusses some preliminaries on multi-coil MRI reconstruction and the approach for searching neighbors that will be used in our algorithm.
Section~\ref{section3} describes the proposed technique and its interpretations.
Section~\ref{section4} presents the experimental setup and results. Section~\ref{section5} provides a discussion of our findings, and in Section~\ref{section6}, we conclude.

\section{Preliminaries} \label{section2}
 
\subsection{Multi-coil MRI Reconstruction}
When an image $\x \in \mathbb{C}^q$ (vectorized) is sufficiently 
sparse in some transform domain and the transform is sufficiently incoherent with the measurement operator, compressed sensing theory~\cite{compress,lustig:07:smt} enables 
accurate image recovery from limited measurements.
The image reconstruction problem in MRI is typically 
formulated as an optimization of a data-fidelity penalty and a regularizer as follows:

\begin{equation}\label{eq:inv_pro}
    \hat{\x}=\underset{\x}{\arg\min} ~\sum_{c=1}^{N_c}\|\mathbf{A}_c \x - \y_c \|^{2}_2 + \lambda  \mathcal{R}(\x),
\end{equation}  
where $\y_c \in \mathbb{C}^p, \ c=1, \ldots, 
N_c,$ represent the acquired k-space measurements 
from $N_c$ coils.
We write the imaging forward operator or measurement operator as $\mathbf{A}_c = \mathbf{M} \mathcal{F} \mathbf{S}_c$,
where $\mathbf{M} \in \{0,1\}^{p\times q}$ is a masking operator that captures the undersampling pattern in k-space,
$\mathcal{F}\in \mathbb{C}^{q\times q}$ is 
the Fourier transform operator (corresponding to densely sampled measurements), and $\mathbf{S}_c \in 
\mathbb{C}^{q\times q}$ is the $c$th 
coil-sensitivity 
matrix (a diagonal matrix). 
Additionally, the regularizer above may include a slew 
of terms capturing the assumed model of the underlying image. 
It enables enforcing desirable properties such as spatial smoothness, image sparsity, or edge preservation in the reconstructed image.
Numerous iterative optimization techniques exist for~\eqref{eq:inv_pro}. 

In MRI, the regularizer can involve $\ell_1$ penalty on wavelet coefficients~\cite{wave} or a total variation penalty~\cite{totalv} or patch-based sparsity in learned dictionaries~\cite{ravishankar2011dlmri} or sparsifying transforms~\cite{wensailukebres19}, or proximity to deep learning-based reconstructions, etc. 
\revise{For example, sparsity w.r.t. a known transform matrix $\mathbf{W}$ is captured by $\mathcal{R}(\x) = \| \mathbf{W} \x\|_{1}$.}

\subsection{Neighbor Search}
Our approach relies on finding images in a data set that are in a sense similar to the one being reconstructed. The similarity may be defined using a metric such as Euclidean distance or other metrics.
Assume we have a 
data set $\left \{ \x_n, \y_n \right 
\}_{n=1}^N$  with $N$ reference or 
ground-truth images 
$\x_n$ and their corresponding k-space measurements $\y_n$ (with multi-coil data), we use the distance metric $d$ to find the $k$ 
nearest neighbors to an (estimated/reconstructed) image $\x$ as follows:

\begin{equation}
\label{eq:bm_formulation}
\begin{aligned}
    \hat{C}_{\x} =\underset{C \in \mathcal{C}, |C|= k}{\arg\min} 
    \sum_{r\in C}d(\x,\x_n),
\end{aligned}
\end{equation}

where $C$ is a set of cardinality $k$ containing indices of feasible neighbors, and $\mathcal{C}$ denotes the set of all such sets with $k$ elements.
Different distance functions could produce a different set of 
similar neighbors, which could then affect the outcome of the 
reconstruction algorithm, as our network modeling is 
dependent on the choice of the local data set.

As a result, we used different 
metrics for evaluating our approach in this work. 
The distances serve as a proxy
for data similarity, with nearby data considered similar 
and distant data considered dissimilar. We used 
the Manhattan distance, Euclidean distance, and normalized cross-correlation as distance metrics as follows.
\begin{align*}
    & d^{L1}(\x,\x_n)  =  \| \x -\x_n\|_1  \\
    & d^{L2}(\x,\x_n)  =  \|\x -\x_n\|_2                   \\
    & d^{NCC}(\x,\x_n)  =  1- \frac{\begin{vmatrix}
\x^{H}\x_n
\end{vmatrix}}{\| \x \|_2 \, \| \x_n\|_2 }.          \label{eq:distance_metric} 
\end{align*}

In all cases, we select 
the top $k$ most similar neighbors (from a set) 
that correspond to the $k$ smallest distances in~\eqref{eq:bm_formulation}.
The indices of the chosen images are in the set $\hat{C}_{\x}$, i.e., they are the minimizer in~\eqref{eq:bm_formulation}.
These neighbors can be used to train 
the local model. 
These are expected to
capture structures most similar to the image being reconstructed, enabling a highly effective reconstruction model to be learned.

\vspace{-0.05in}
\section{Proposed LONDN-MRI Algorithm} \label{section3}
Our primary objective is to learn an adaptive neural network for MRI reconstruction, 
in which the model's free parameters are 
fitted using training data that are similar in a sense to the current scan. 
We emphasize that the proposed 
model is local in the sense that it changes in
response to the input. 
The 
advantage of the proposed method is that the model is fit for every scan and can thus be adaptive to the scan, readily handling changes in sampling masks, for example. 

The algorithm begins by obtaining an initial estimate of the underlying image, denoted $\x^0$, from undersampled measurements $\y$.  
Our proposed strategy then 
alternates between computing the closest neighbors 
to the reconstruction
in the training set and performing 
CNN-based supervised learning on the estimated local 
dataset. 
During supervised learning, the network weights could be randomly initialized or could be warm started with the weights of a pre-trained (e.g., state-of-the-art) network. In the latter case, the pre-trained network would adapt to the features of images similar to the one being reconstructed (akin to transfer learning~\cite{transfer_learning_MRI}).

In each iteration, the nearest ground truth images in the training
set are computed in relation to the reconstruction 
(estimate) predicted by the locally learned network, 
except in the first iteration, when the nearest neighbors 
are computed in relation to the (typically highly aliased)
initial $\x^{0}$ (we used corresponding aliased images in the dataset for computing distances in the first iteration).  
In practice, pairwise distances to even a 
large number of 
images can be computed very efficiently (in parallel), 
after which the local network can be 
rapidly learned on a 
small set of neighbors (typically a 
shallow network or with early stopping). 
The network weights for 
deep reconstruction are constantly updated to map the 
initial images for the local data set to the target 
(ground truth) versions.

To demonstrate our approach, we used 
the state-of-the-art deep CNN 
reconstruction model~MoDL~\cite{modl}, 
which is trained locally in our scheme. 
Additionally, we trained it globally, i.e., once on a 
larger dataset, 
in 
order to compare it to our on-the-fly 
neighborhood-based 
learning scheme. 
For completeness, we briefly recap the MoDL
scheme in the 
following and discuss its local 
training within our framework. 
MoDL is similar to the plug-and-play approach, except that
instead of pre-trained denoiser networks, end-to-end 
training is used to learn the shared network
weights across iterations in the architecture.

\subsection{Network Model and Training}
The proposed approach is compatible with any 
network architecture. We use MoDL, which has shown promise for MR image reconstruction, and combines a 
denoising network with a data consistency (DC) module in each iteration of an unrolled architecture. MoDL unrolls alternating minimization for the following problem:
\begin{equation}
    \text{L}_a(\z,\x) := \nu \sum_{c=1}^{N_c} \|\A_c \x - \y_c\|_2^2 +  \mathcal{R}(\z)+\mu\|\x - \z\|_{2}^2.
    \label{eq:altmin}
\end{equation}
We denote the initial image in the process as  $\x^{0}$, 
$\nu \geq 0$ weights the data-consistency term above, and $\mu \geq 0$ weights the proximity of $\x$ to $\z$. 
By decomposing the optimization into two subproblems over $\z$ and $\x$, the explicit regularizer-based update for $\z$ can be solved by replacing it with a  CNN-based denoiser ($D_{\theta} (\cdot)$), and the denoised estimate is then used to update $\x$. The $\x$ update in the MoDL scheme involves the data-consistency term and is performed using Conjugate Gradient (CG) descent.
Thus, $\z$ is obtained as the output from a CNN-based denoiser ($D_{\theta}$) and $\x$ is updated by CG. 
This alternating scheme is repeated $L$ times (unrolling), with the initial input image $\x^{0}$ being passed through $L$ blocks of denoising CNN + CG updates.
Now, if  $\Sbs{l}{\theta}(.)$ is the function capturing the $l$th iteration of the algorithm, 
then the MoDL output for the $l$th block
is given as
\begin{equation}
\begin{split}
    \label{modleqn1}
    &\x^{l+1} = \Sbs{l}{\theta}(\x^l) =
    {\Sb}\big(\x^l,\theta,\nu_l, \{\A_c,\y_c\}_{c=1}^{N_c} \big)
    , \, \text{and}\\
    &\Sb\big(\bar{\x},\theta,\nu,\{\A_c,\y_c\}_{c=1}^{N_c}\big)
    \defequ
    \\&
    \argmin{\x}~ \nu \sum_{c=1}^{N_c} \|\A_c \x - \y_c\|_2^2
    + \|\x-\D_\theta(\bar{\x})\|_2^2.
\end{split}
\end{equation}
After $L$ iterations, the final output is
\begin{gather}\label{eq:supervised_loss}
    \x_{\text{supervised}} = \x^L = \paren{\bigcomp_{l=0}^{L-1}\Sbs{l}{\theta}}(\x^0)\defequ \scrM_{\theta}(\x^0),
\end{gather}
where $\scrM_{\theta}$ denotes the end-to-end mapping, $\bigcomp_{i=0}^{L-1} f^i$ represents 
the composition of $L$ functions 
$f^{L-1}\circ f^{L-2}\circ\ldots\circ f^0$,
and $\x^{0}$ is the  initial image.
The weights of the denoiser $\D_\theta$
are shared across the $L$ blocks.
The network parameters $\theta$ are 
learned in a supervised manner
so that $\x_\text{supervised}$ matches 
known ground truths
(in mean squared error or other 
metric) on a (large/global or local) training 
set. 
This involves the following optimization for training:
\revise{
\begin{equation}
\begin{split}
    \hat{\theta} & =
    \argmin{\theta}~ \sum_{n \in S}C_{\beta}(\scrM_{\theta}(\x_n^{0}); \x_{\text{n}})
    \nonumber\\ 
    & = \argmin{\theta}
    \sum_{n \in S} \big( \big \|\x_n - \scrM_{\theta}(\x_n^{0}) \big\|_2^2
 )
,
\end{split}
\label{eq:suptrn_cost}
\end{equation}}
where
$n$ indexes the samples from the data 
set used for training, with 
\revise{$\x_{\text{n}}$} denoting the $n$th 
target (or ground truth) image 
reconstructed from fully-sampled 
k-space measurements and \revise{$\x_n^{0}$} 
denotes the initial image estimate from
undersampled measurements.
The cost $C_{\beta}(\hat{\x}_n; \x_n)$
denotes the training loss.
The main difference between a globally 
learned and locally learned network is the choice of
the set $S$ of training indices. 
For the proposed local approach, we fit
the network based on the $k$ training 
samples closest to the current test 
image estimate, whereas the 
conventional (or global) training would fit 
networks to a large dataset.
The initial image estimate $\x_n^{0}$ 
is obtained from the undersampled 
measurements $\y_n$
using a simple analytical 
reconstruction scheme such as
applying the adjoint of the forward 
model to the measurements.

\begin{figure}
\vspace{-0.05in}
\centering
\setlength{\tabcolsep}{0.4cm}
\begin{tabular}{cc}
\includegraphics[width=1.0\linewidth]{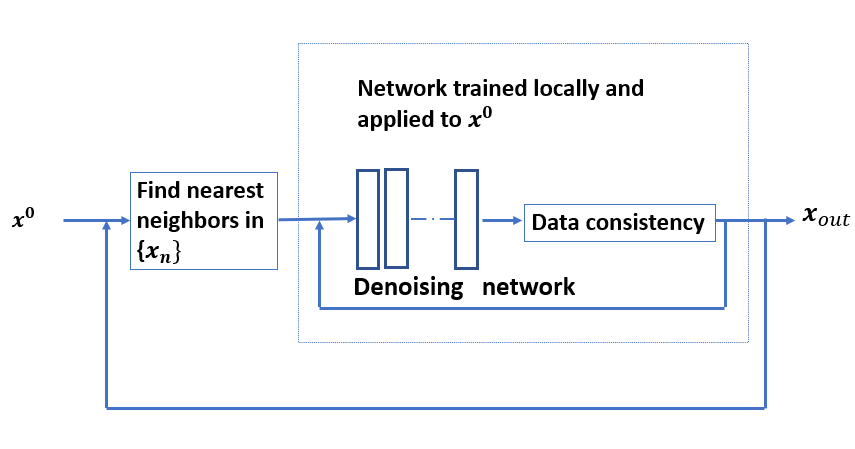}
\end{tabular}
    \caption{Flowchart of the proposed LONDN-MRI scheme with a specific unrolled reconstruction network. The denoising network could be for example a U-Net or the recent DIDN.}
    \label{algorithmflowchart}
    
\end{figure}

In each iteration, the network is 
updated (Fig.~\ref{algorithmflowchart}), 
and the initial estimate of the 
underlying unknown image is passed 
through the network to obtain a new 
estimate. In 
Fig.~\ref{algorithmflowchart}, we 
illustrate the iterative process of neighbor
fine-tuning and local network updating. 
Local learning may have the advantage 
of accommodating changes in 
experimental conditions (e.g., 
undersampling pattern) at test time, 
provided that such modified 
measurements and initial images for the small local
training set can be easily simulated 
from the existing $\x_n$ or $\y_n$.
Our overall algorithm is also summarized in Algorithm~\ref{alg::ADMM_anisoTV}.


\begin{algorithm}
\caption{LONDN-MRI Algorithm}
\label{alg::ADMM_anisoTV}
\begin{algorithmic}[1]
\Require Initial image $\x^{0}$, number of neighbors $k$, k-space undersampling 
mask $\mathbf{M}$, regularization parameters 
$\nu$ and $\mu$, number of training epochs $T$, number of iterations of alternating algorithm $S$. 
\State Initialize reconstruction network parameters $\theta$ with pre-learned network weights $\hat{\theta}$ or randomly initialized weights. Set $\x = \x^{0}$.
\For{Iteration $<$ maximal iteration $S$}
    \State Compute the set of $k$ similar neighbors $\hat{C}_{\x}$ to the current reconstruction estimate $\x$ using metric $d$. 
    \For{epoch $<$ maximal number $T$}
    \State For each batch of neighbor data, compute the gradient of the training loss with respect to the network parameters $\theta$ and perform one update step on $\theta$.
    \EndFor
    \State Update  $\x$ $\leftarrow \scrM_{\theta}(\x^{0})$
\EndFor     

\State \textbf{return} reconstruction $\x$ and learned net. parameters $\theta$.
\label{algorithm_londn}
\end{algorithmic}
\end{algorithm}

\subsection{Regularization}
In order to avoid over-fitting when training networks on small sets, we also adopted regularization of weights during training as follows:
\revise{
\begin{equation}
\begin{split}
    \hat{\theta} = \argmin{\theta}
    \sum_{n \in S} \big \|\x_n - \scrM_{\theta}(\x_n^{0}) \big\|_2^2 + \lambda \, \mathcal{R}(\theta),
\end{split}
\label{eq:suptrn_cost_regularization}
\end{equation}}
where $\mathcal{R}(\cdot)$ denotes the regularization term on network weights. We primarily used the $\ell_1$ norm regularizer to enforce sparsity of the network weights to learn simpler models.
We observed that regularizing the
local model enables it to converge more easily, and shrinks 
weights for less important or noisy features to zero.
We provide more discussion in the experiments section.
\subsection{Connections to Bilevel Optimization}
The alternating algorithm for training involving a neighbor search step and a local network update step could be viewed as a heuristic algorithm for the following bilevel optimization problem:
\revise{
\begin{align}
 \nonumber &   \min_{C \in \mathcal{C}, \, |C|=k} \sum_{i\in C} || f_{{\theta}(C)}(\y) - \x_{i}||_2^{2},\\
 \text{s.t.} & \;\;\;   \theta(C) = \arg \min_{\theta} \sum_{i\in C} || \x_{i} - f_{\theta}(\y_i)||_2^{2}.
\label{eq:bilevel_problem}
\end{align}}

Here, $ f_{{\theta}(C)}$ denotes a deep neural network learned on a subset $C$ of a data set that maps the current k-space measurements $\y$ to a reconstruction. The network is akin to $\scrM_{\theta}(\x^{0})$ shown earlier~\eqref{eq:suptrn_cost}, but with $\x^{0}$ assumed to be generated from $\y$ (e.g., via the well-known sum of squares of coil-wise inverse Fourier transforms, or via SENSE reconstruction, etc.).
Problem~\ref{eq:bilevel_problem} aims to find the best neighborhood or cluster among the training data, where the reconstructed image belongs (with closest distances to neighbors -- we assumed Euclidean distance here), with the network weights for reconstruction estimated on the data in that cluster.
Problem~\ref{eq:bilevel_problem} is a bilevel optimization problem with the cluster optimization forming the upper level cost and network optimization forming the lower level cost. Bilevel problems are known to be quite challenging~\cite{crockett2021bilevel,9747201}.
It is also a combinatorial problem because we would have to sweep through all possible choices of clusters of $k$ training samples with reconstruction networks trained in each such cluster, to determine the best cluster choice.

The proposed algorithm is akin to optimizing the bilevel problem by optimizing for the network weights $\theta$ with the clustering $C$ fixed (the lower level problem) and then optimizing for the clustering $C$ (upper level minimization) with the network weights fixed. This is a heuristic because the optimized variables in each step are related, however, such an approach has been used in prior work~\cite{super} and shown to be approximately empirically convergent for the bilevel cost.
In this work, we performed an empirical evaluation of convergence in the experiments section, where the alternating algorithm is shown to reduce the upper-level cost in~\eqref{eq:bilevel_problem}.

\begin{figure*}
\centering
\setlength{\tabcolsep}{0.5cm}
\includegraphics[width=0.24\linewidth]{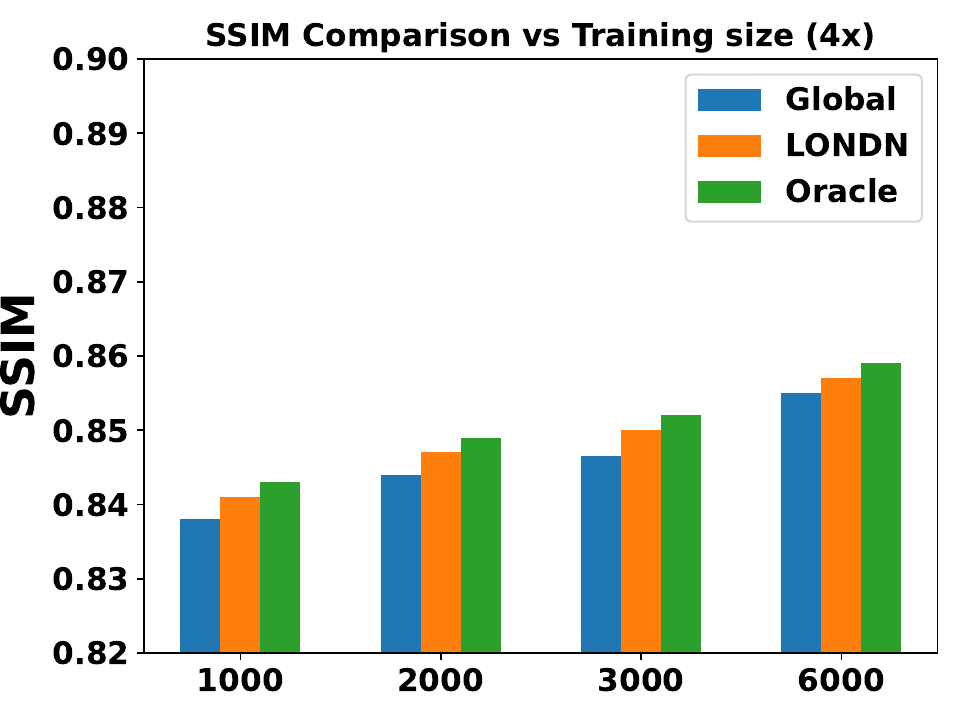}
\includegraphics[width=0.24\linewidth]{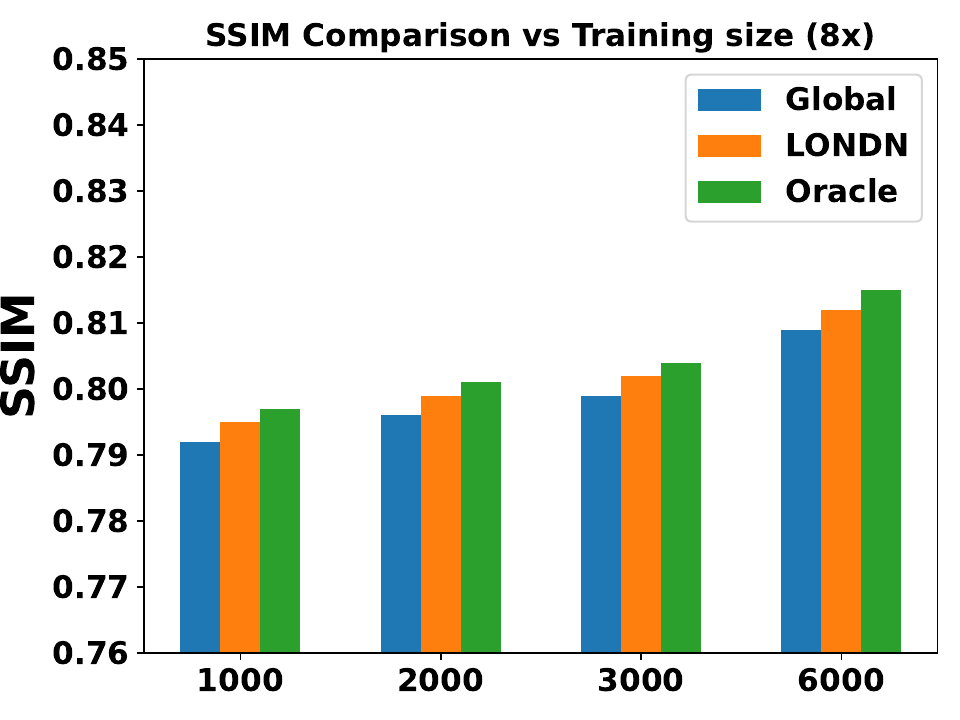}
\includegraphics[width=0.24\linewidth]{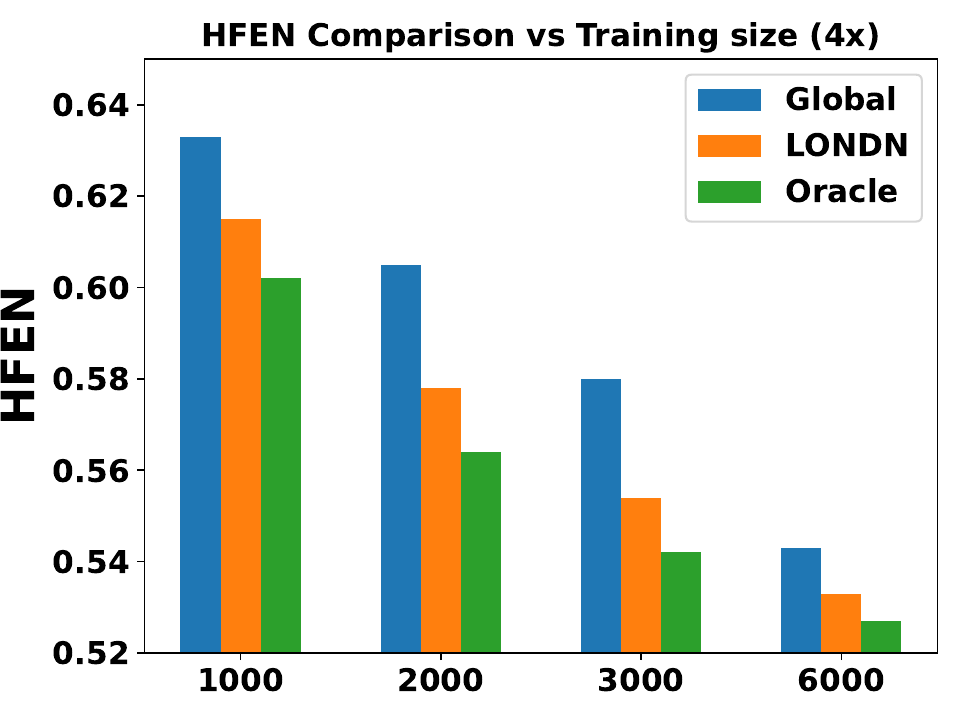}
\includegraphics[width=0.24\linewidth]{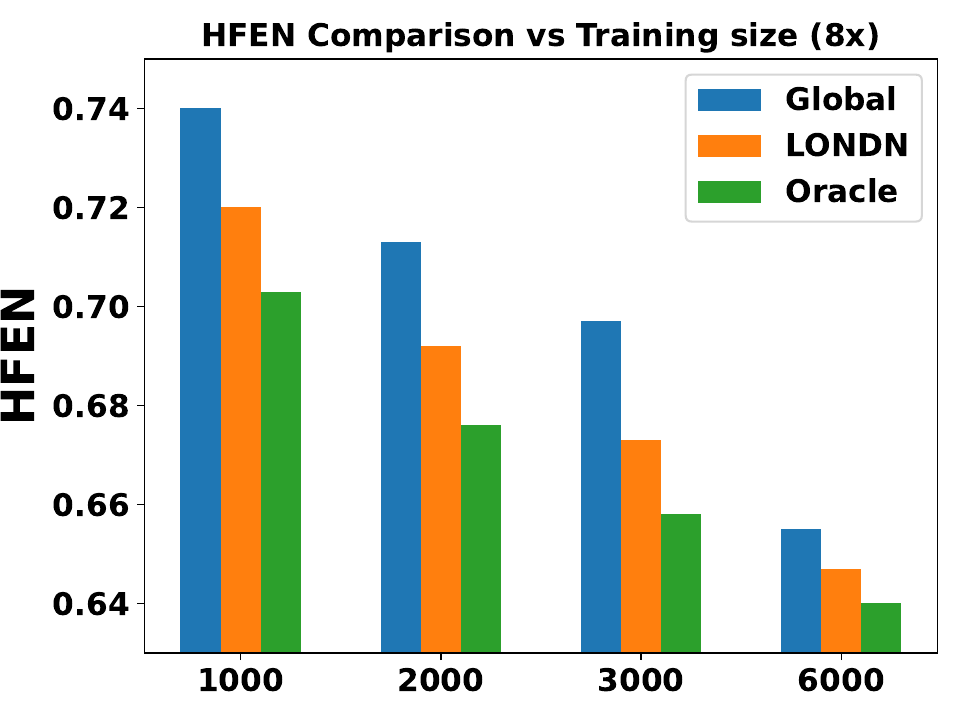}
\caption{ Comparison of MoDL with UNet denoiser trained globally vs. using the proposed LONDN-MRI scheme (1 iteration). Reconstruction metrics are shown across training set sizes at 4x and 8x undersampling.}
\label{PSNR_SSIM_HFEN}
\end{figure*}

\section{Experiments} \label{section4}

\SR{We first present the overall experimental setup in Section~\ref{setting}. Key results and comparisons are presented in Section~\ref{result}. The intricacies and behavior of LONDN-MRI are analyzed in Section~\ref{behavior} and its generalizability is investigated in Section~\ref{Generalizability}.}

\subsection{Experimental Setup}\label{setting}
\noindent \textbf{Datasets \& Models:} \SR{We evaluated the effectiveness of the proposed LONDN-MRI reconstruction method on multiple datasets: the multi-coil fastMRI knee and brain datasets~\cite{zbontar2019fastmri,knoll2020fastmri}, the fastMRI+ dataset\footnote{{\url{https://github.com/microsoft/fastmri-plus/tree/main}}} (which is just an annotated version of fastMRI indicating pathologies), and the Stanford 2D FSE~\cite{FSE} dataset. The results obtained on the fastMRI knee dataset and the Stanford FSE data are described in Section~\ref{result}. The fastMRI brain and fastMRI+ data are used in the studies in Section~\ref{Generalizability}. For training, we randomly selected a subset of $3000$ images from the fastMRI knee and brain datasets and the same for the fastMRI+ case. We used $2000$ training images for the smaller Stanford FSE dataset. We used $15$ or $20$ images for testing in different scenarios, which were randomly chosen. 
In some experiments, we evaluated the effect of training set size, where we worked with fewer or more images in the training set.}

Coil 
sensitivity maps for model-based reconstruction were generated for each scan 
using the BART toolbox~\cite{martin_uecker_2018_1215477}.
\SR{We tested obtaining these using either the fully-sampled k-space 
or only center of k-space data 
and noticed very little difference in reconstruction quality between the two approaches.}

Since the proposed LONDN-MRI framework is quite general and can be combined with any supervised deep learning based reconstruction approach, we chose the recent popular model-based deep learning (MoDL) reconstruction network and compared globally (over large set of training samples) and locally (over very small matched set of samples) learned versions of the model for different choices of deep denoisers in the network
\footnote{
See
\url{https://github.com/sjames40/Multi_coil_local_model}
for our code in PyTorch.
}.
We performed reconstructions at fourfold or 4x
acceleration (25.0\% sampling) as well as at eightfold or 8x acceleration (12.5\% sampling) of the k-space acquisition. In all cases, variable density
1D random Cartesian (phase-encode) undersampling of k-space was performed. 
The initial image estimates for MoDL were obtained by applying the adjoint of the measurement operator to the subsampled k-space data,
and were then used to train both local and global versions of MoDL networks.
In our local versions (LONDN-MRI), we used 30 images for training (searched from e.g., 3000 images). while the global versions used the full subset of training images.

\noindent \textbf{Network Architectures \& Training:} We trained two types of MoDL models at 4x and 8x k-space undersampling, respectively.
One used the well-known UNet denoiser, with a two-channel input and two-channel output, where the real and imaginary parts of an image are separated into two channels. The network weights during training were initialized randomly (normally distributed). The 
ADAM optimizer was 
utilized for training the network weights. For LONDN-MRI, we used an initial learning rate of $6 
\times 10^{-5}$ with a multi-step learning rate scheduler, which decreases the learning rate at $100$ and $150$ epochs with learning rate decay $0.65$.
\revise{For training globally, we used an initial learning rate of $1 
\times 10^{-4}$ with $150$ epochs of training and a multi-step learning rate scheduler that decreased the learning rate at $50$ and $100$ epochs with learning rate decay $0.6$.}
For LONDN-MRI, MoDL with $5$ 
iterations was used
with a shallow UNet that had $2$ layers in the encoder and
decoder, respectively. 
We used a shallow network with dropout for the local model to avoid over-fitting to the very small training set. 

For the MoDL network trained globally (on large dataset) for making comparisons with, we utilized $4$ layers in the decoder and encoder in UNet and $6$ MoDL blocks. 
We used a batch size of $2$ during training for both the global and local cases. Furthermore, for the data-consistency term, we used a tolerance of $10^{-5}$ in CG and a $\mu/\nu$ ratio of 0.1. \revise{Also, we chose the regularization weight $\lambda$ as $10^{-9}$ for LONDN-MRI, unless specified otherwise.}


For the second MoDL architecture, we used the recent state-of-the-art denoising network DIDN~\cite{DIDN_ori,blipstmi2021}.
\SR{Due to the high complexity of the DIDN 
network, we first pre-trained it on the larger (global) dataset \revise{(learning rate, etc., similar to the UNet case)} before adapting the weights within LONDN-MRI for each scan.
This is an alternative to constructing shallower versions of a network for local adaptation.}

The 
ADAM optimizer was utilized for training, with a 
learning rate of $5\times
10^{-5}$ in LONDN-MRI. We used $6$ iterations of MoDL with the 
DIDN denoiser for which we used $3$ down-up blocks (DUBs). The number of epochs for training was $30$ in LONDN-MRI.
The remaining training 
parameters were chosen similarly as in the previous UNet-based case.
\SR{Using a pre-trained state-of-the-art denoiser allows the local adaptation to converge faster.}

\noindent \textbf{Comparison to Scan-adaptive Methods:}
\SR{We compared the performance of our schemes to recent related scan-specific methods such as deep image prior (DIP)~\cite{DIP_MRI}(using the public package\footnote{\url{https://github.com/MLI-lab/ConvDecoder}} but additionally incorporating coil sensitivity maps), RAKI~\cite{Raki} (using publicly available code\footnote{\url{https://github.com/geopi1/DeepMRI}}), SOUP-DIL~\cite{SOUP-DIL} (code extracted from publicly available package\footnote{\url{https://github.com/JeffFessler/BLIPSrecon/}}), and LORAKI\cite{loraki} (modified from RAKI code). In our experiments, we used parameters specified in the authors' original implementations, which we observed worked well.}

\begin{figure}[h!]
\begin{center}
    \begin{tabular}{cc}
        \includegraphics[height=1.0in]{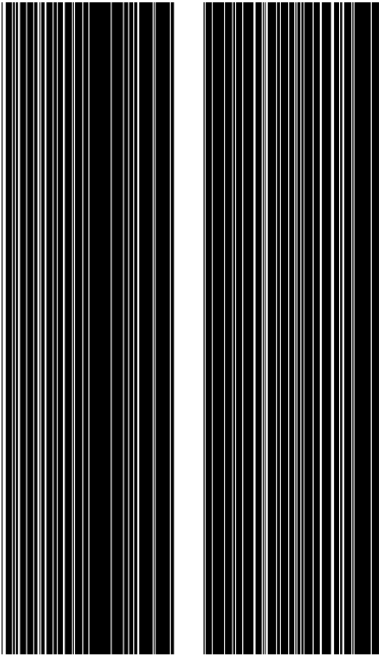}&
        \includegraphics[height=1.0in]{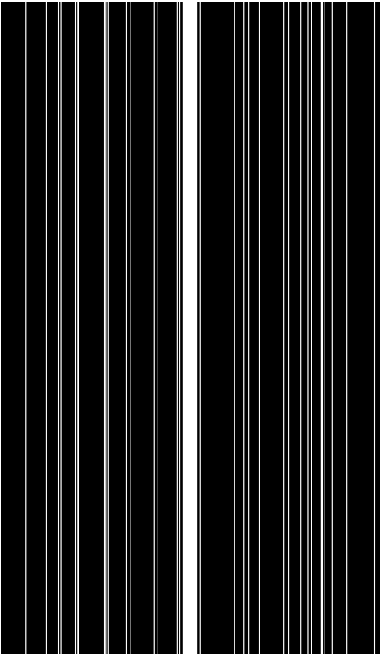}\\
        (a) & (b)  \\
\end{tabular}
\caption{Undersampling masks used in our experiments:
(a) fourfold undersampled 1D Cartesian phase-encoded; and
(b) eightfold undersampled 1D Cartesian phase-encoded. The masks were zero-padded for slightly larger images.
}
\label{fig:usml_msk}
\end{center}
\end{figure}

\noindent \textbf{Sampling Masks \& Performance Metrics:} We used binary masks for fourfold and eightfold Cartesian undersampling of k-space. Fig.~\ref{fig:usml_msk} shows the sampling masks primarily used in our experiments that include a fully-sampled central region (with $31$ central lines at 4x acceleration and $15$ central lines at 8x acceleration) and the remaining phase encode lines were sampled uniformly at random. 

For the performance metrics, we used three common metrics to quantify the reconstruction quality of different methods. 
These were the peak signal-to-noise ratio (PSNR) in decibels (dB), structural similarity index (SSIM)~\cite{wang2004image}, and the high frequency error norm (HFEN)~\cite{ravishankar2011dlmri}, which were computed between the reconstruction and the ground truth obtained from fully-sampled k-space data.
The HFEN was computed from the $\ell_2$ norm of the difference between Laplacian of Gaussian (LoG) filtered reconstructed and ground truth images. This was normalized by the $\ell_2$ norm of the LoG filtered ground truth.

\subsection{Results and Comparisons}
\label{result}
\noindent \textbf{Results for the UNet-based Reconstructor:} 
Table~\ref{table:PSNR_comparison_unet}
compares the average PSNR values for reconstruction
over the fastMRI knee testing set at both 4x and 8x undersampling.
We varied the number of images in the training set for a more comprehensive study. We compare learning networks over a small set of similar images to learning networks over the larger datasets (global), as well as to an oracle LONDN scheme, where the neighbors in the training set were computed based on each ground truth test image.
The oracle scheme would ideally provide an upper bound on the performance of the iterative LONDN-MRI scheme. \SD{Moreover, LONDN-MRI 
outperforms DIP, RAKI, SOUP-DIL, and LORAKI with  U-Net (Table~\ref{table:PSNR_comparison_unet}).
Note that DIP, RAKI, SOUP-DIL, and LORAKI do not use information beyond the test scan (scan-adaptive). Later, we show how LONDN-MRI performs when the overall dataset it uses is very limited.}

When varying the size of the training set, the global approach was trained on the full set each time, whereas the local approach performed training on small subsets of $30$ training pairs selected from the larger datasets.
The iterations of the LONDN-MRI scheme quickly improve reconstruction performance, and even with only 2 LONDN-MRI alternations, the PSNR values begin approaching the oracle setting.The LONDN schemes (oracle or iterative) consistently outperform the globally trained networks across the different training set sizes considered.
\SR{We note that the results for the globally trained model with many ($6000$) training scans match closely the LONDN-MRI results, when LONDN-MRI uses a smaller overall training set ($3000$ scans) for neighbor search. This illustrates the potential of our approach with limited training data, when compared with models trained on larger sets.}

Figure~\ref{PSNR_SSIM_HFEN} compares the SSIM and HFEN reconstruction metrics using bar graphs, where a similar trend is observed as with PSNR.

\begin{figure*}
\centering
\begin{tabular}[b]{ccccc}
    \textbf{Ground Truth}
    & \textbf{Global} 
    & \textbf{LONDN-MRI}
    & \textbf{LONDN-MRI} 
    & \textbf{Oracle} \\
     \textbf{}
    & \textbf{} 
    & \textbf{(1 iteration)}
    & \textbf{(2 iterations)} 
    & \textbf{}\\
    
    \includegraphics[width=.13\linewidth,valign=t]{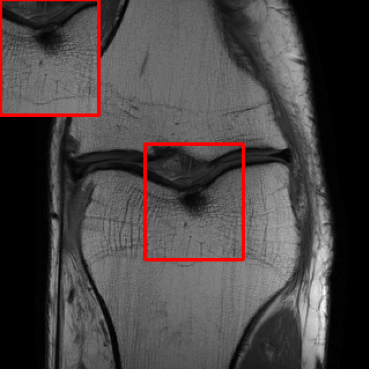} &
    \includegraphics[width=.13\linewidth,valign=t]{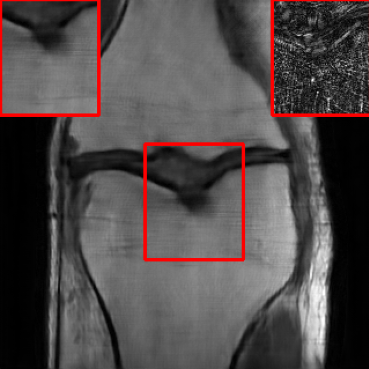} &
    \includegraphics[width=.13\linewidth,valign=t]{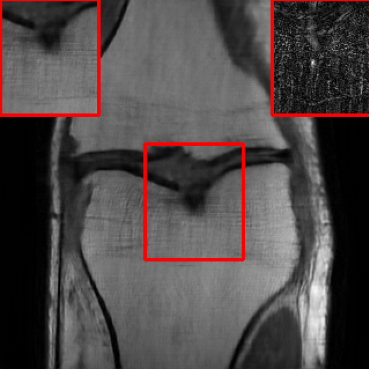} &
    \includegraphics[width=.13\linewidth,valign=t]{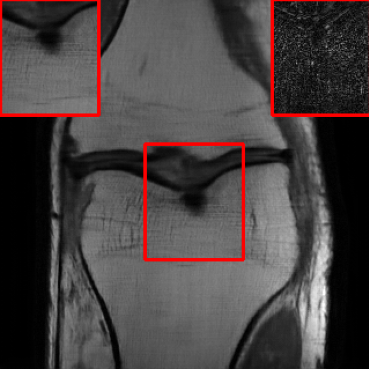} &
    \includegraphics[width=.13\linewidth,valign=t]{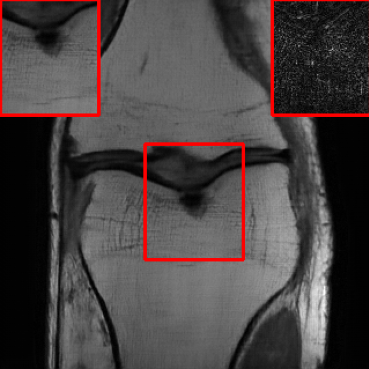} \\
     \scriptsize{ PSNR = $\infty$ dB}&  \scriptsize{ PSNR = 29.26 dB} & (d)  \scriptsize{PSNR = 29.50 dB} &  \scriptsize{ PSNR = 29.68 dB} & (f)  \scriptsize{PSNR = 29.72 dB}\\
    \textbf{Initial} 
    & \textbf{DIP} 
    & \textbf{SOUP-DIL}
    & \textbf{RAKI} 
    & \textbf{LORAKI} \\
    \includegraphics[width=.13\linewidth,valign=t]{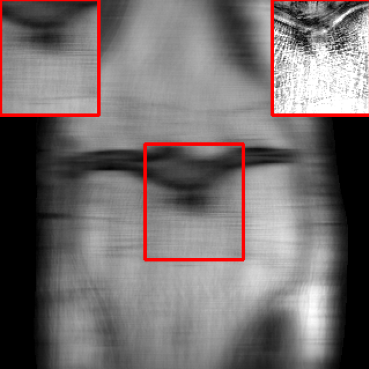} &
    \includegraphics[width=.13\linewidth,valign=t]{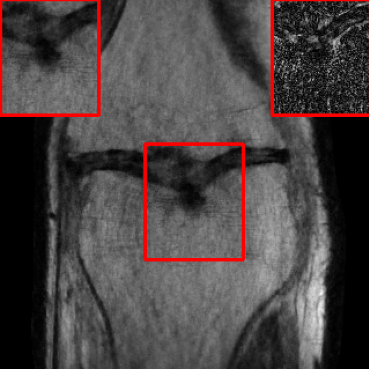} &
    \includegraphics[width=.13\linewidth,valign=t]{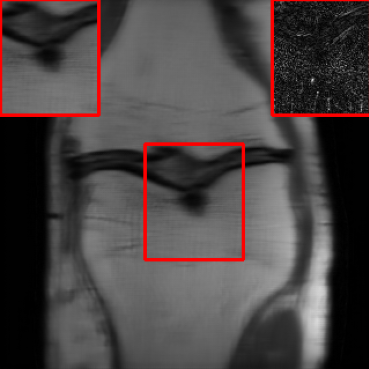} &
    \includegraphics[width=.13\linewidth,valign=t]{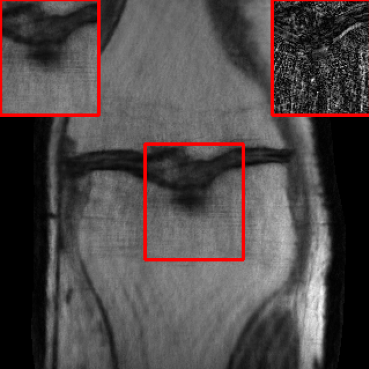} &
    \includegraphics[width=.13\linewidth,valign=t]{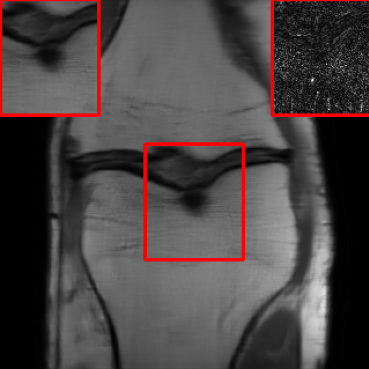} \\
    \scriptsize{PSNR = 16.43 dB} & \scriptsize{ PSNR = 27.50 dB} & (d)  \scriptsize{PSNR = 28.16 dB} &  \scriptsize{ PSNR = 28.32 dB} & (f)  \scriptsize{PSNR = 28.72 dB}
\end{tabular}%
\caption{\SR{Comparison of image reconstructions with different methods at 8x undersampling. The global and LONDN-MRI methods use the MoDL architecture with UNet denoiser with 1000 training images. The inset panel on the top left in each image corresponds to a section of interest in the image (shown by the red bounding box), while the inset panel on the top right corresponds to the error map with respect to the ground truth.}}
\label{fig:denoised_imgs_zoomed1}
\end{figure*}

\begin{figure*}
\centering
\begin{tabular}[b]{ccccc}
    \textbf{Ground Truth}
    & \textbf{Global} 
    & \textbf{LONDN-MRI}
    & \textbf{LONDN-MRI} 
    & \textbf{Oracle} \\
     \textbf{}
    & \textbf{} 
    & \textbf{(1 iteration)}
    & \textbf{(2 iterations)} 
    & \textbf{}\\
    \includegraphics[width=.13\linewidth,valign=t]{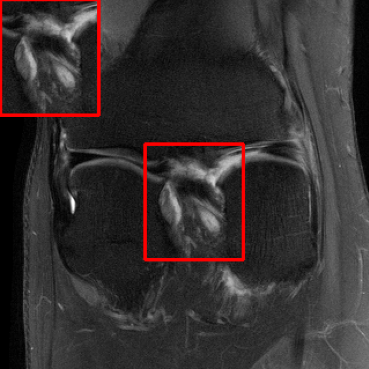} &
    \includegraphics[width=.13\linewidth,valign=t]{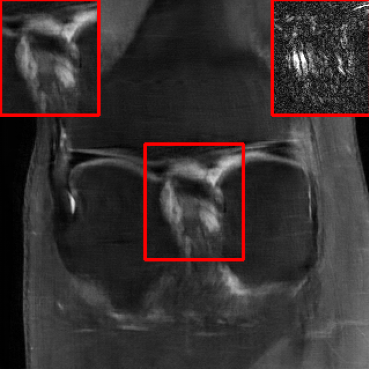} &
    \includegraphics[width=.13\linewidth,valign=t]{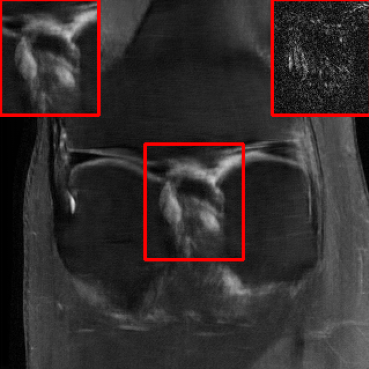} &
    \includegraphics[width=.13\linewidth,valign=t]{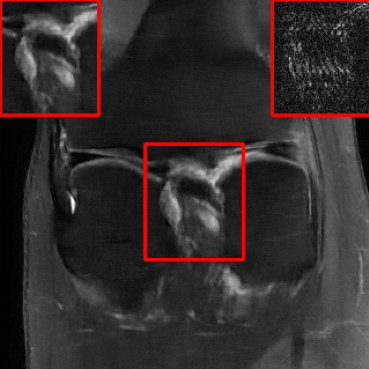} &
    \includegraphics[width=.13\linewidth,valign=t]{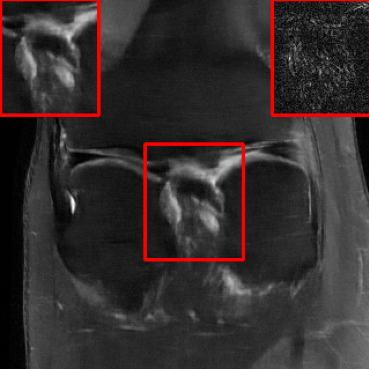} \\
    \scriptsize{PSNR = $\infty$ dB} &   \scriptsize{PSNR = 32.78 dB} &  
    \scriptsize{PSNR = 33.16 dB} &
    \scriptsize{PSNR = 33.25 dB} &  
    \scriptsize{ PSNR = 33.30 dB}\\
    \textbf{Initial} 
    & \textbf{DIP} 
    & \textbf{SOUP-DIL}
    & \textbf{RAKI} 
    & \textbf{LORAKI} \\
    \includegraphics[width=.13\linewidth,valign=t]{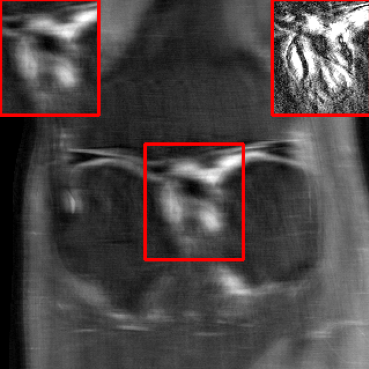} &
    \includegraphics[width=.13\linewidth,valign=t]{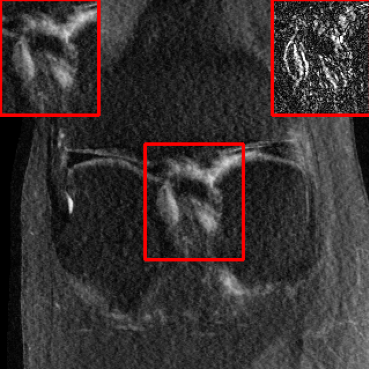} &
    \includegraphics[width=.13\linewidth,valign=t]{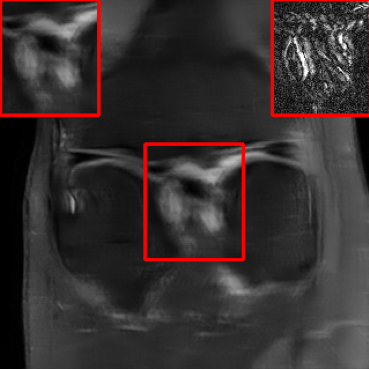} &
    \includegraphics[width=.13\linewidth,valign=t]{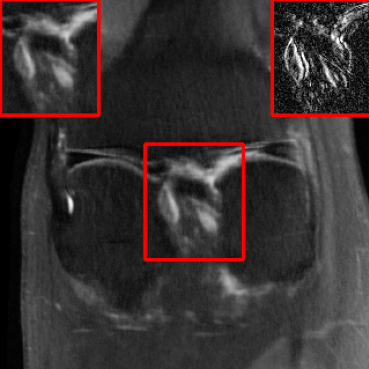} &
    \includegraphics[width=.13\linewidth,valign=t]{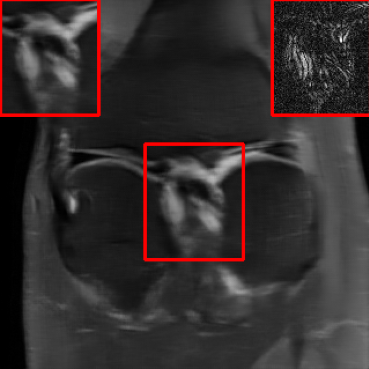} \\
    \scriptsize{PSNR = 21.23 dB} &  \scriptsize{ PSNR = 30.18 dB} &  \scriptsize{PSNR = 30.66 dB} &  \scriptsize{PSNR = 31.26 dB} &   \scriptsize{ PSNR = 31.67 dB}
\end{tabular}

\caption{\SR{Same comparisons/setup as Fig.~\ref{fig:denoised_imgs_zoomed1}, but at 4x undersampling. The supervised methods used MoDL architecture with UNet denoiser ($3000$ training images).}}
\label{fig:denoised_imgs_zoomed2}
\end{figure*}

Figs.~\ref{fig:denoised_imgs_zoomed1} and~\ref{fig:denoised_imgs_zoomed2} show images reconstructed by different methods at 8x and 4x undersampling, respectively. 

The LONDN-MRI reconstructions (either iterative or oracle) show fewer artifacts, sharper features, and fewer errors than the global MoDL and initial aliased reconstructions.
The iterative LONDN-MRI results are also quite close to the oracle result. 

\begin{table}[htp!]
\centering
\resizebox{0.5\textwidth}{!}{
\addtolength{\tabcolsep}{-2.1pt}
\begin{tabular}{cccccccccc}
\toprule
\multicolumn{1}{c}{Ax}&
\multicolumn{1}{c}{Data}&
\multicolumn{1}{c}{Global}& 
\multicolumn{1}{c}{LONDN-MRI}&
\multicolumn{1}{c}{LONDN-MRI} & 
\multicolumn{1}{c}{Oracle} &\multicolumn{1}{c}{DIP}&
\multicolumn{1}{c}{RAKI}&\multicolumn{1}{c}{LORAKI}&\multicolumn{1}{c}{SOUP} \\
\multicolumn{1}{c}{}& 
\multicolumn{1}{c}{size}& 
\multicolumn{1}{c}{}& 
\multicolumn{1}{c}{(1 iteration)}&
\multicolumn{1}{c}{(2 iterations)} & 
\multicolumn{1}{c}{}
&\multicolumn{1}{c}{}&\multicolumn{1}{c}{}&

\multicolumn{1}{c}{}&\multicolumn{1}{c}{DIL}\\
\hline
\multirow{3}{*}{$\mathrm{4x}$} &1000& 32.63  &32.78  & \textbf{32.87}  & 32.99 & \multirow{3}{*}{30.1}&\multirow{3}{*}{30.25} &\multirow{3}{*}{31.35} &\multirow{3}{*}{30.97}\\
 &2000& 33.00  &33.28  &  \textbf{33.31} & 33.35\\
 &3000& 33.17  & 33.46 & \textbf{33.51}  &33.54 \\
  &6000& 33.48  & 33.58 & \textbf{33.65}  &33.69 \\
 \hline
\multirow{3}{*}{$\mathrm{8x}$} &1000 & 29.78  & 30.15 & \textbf{30.26} &30.34 & \multirow{3}{*}{28.9}&\multirow{3}{*}{29.01} &\multirow{3}{*}{29.71} &\multirow{3}{*}{29.47}\\
 &2000 &  30.21 &30.53  &\textbf{30.58}  & 30.64 \\
 &3000& 30.47  & 30.76 &\textbf{30.80}  & 30.85 \\
  &6000& 30.78  & 30.94 &\textbf{31.04}  & 31.09 \\
\bottomrule \vspace{0.01in}
\end{tabular}}
\caption{\SR{Average reconstruction PSNRs (in dB) for 15 images at 4x and 8x k-space undersampling. The proposed LONDN-MRI (with 1 or 2 alternations) is compared to training a global reconstructor for different training set sizes and another scan based method. We also compare to an oracle local reconstructor, where neighbors are found with respect to known ground truth test images. }}
\label{table:PSNR_comparison_unet}
\end{table}

\begin{figure*}[h]
\centering
\begin{tabular}{cccc}
    \textbf{Ground Truth} & \textbf{Global} & \textbf{LONDN-MRI} & \textbf{Oracle} \\
    \textbf{} & \textbf{} & \textbf{(1 iteration)} & \textbf{} \\
    \includegraphics[width=.13\linewidth,valign=t]{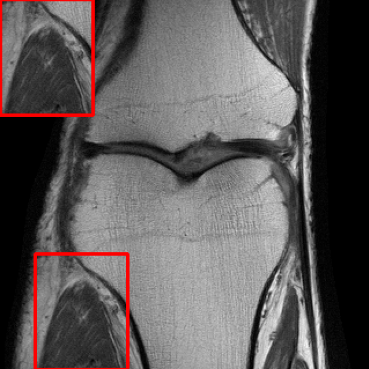} &
    \includegraphics[width=.13\linewidth,valign=t]{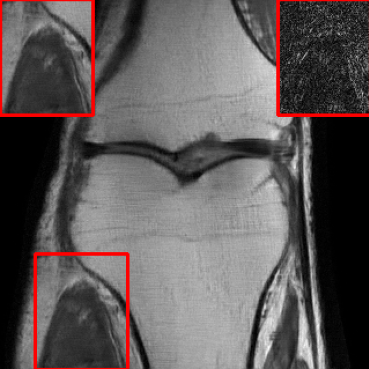} &
    \includegraphics[width=.13\linewidth,valign=t]{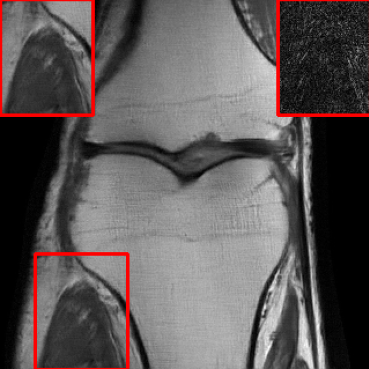} &
    \includegraphics[width=.13\linewidth,valign=t]{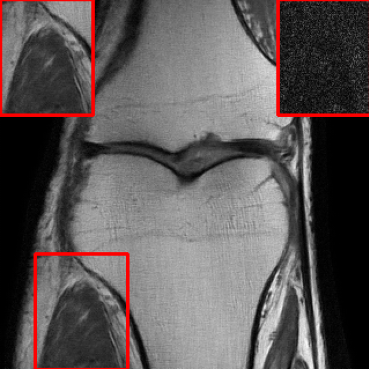} \\
     \scriptsize{PSNR = $\infty$ dB} & 
    \scriptsize{PSNR = 34.15 dB} &
   \scriptsize{PSNR = 34.46 dB} &
  \scriptsize{PSNR = 34.54 dB} \\
\end{tabular}
\caption{Comparison of image reconstructions at 4x undersampling for the MoDL network with DIDN denoiser and $3000$ training images, when compared to LONDN-MRI.
A region of interest and its error are also shown.}
\label{fig:denoised_imgs_zoomed3}
\end{figure*}

\noindent \textbf{Results for the DIDN-based Reconstructor:} \SR{To demonstrate adaptability to different network architectures, Table~\ref{table:PSNR_comparison_DIDN} compares reconstruction performance on the test set with the DIDN denoiser-based MoDL architecture.} Average PSNR values with LONDN-MRI are compared to those with networks trained globally at different training set sizes. We ran only $1$ iteration of LONDN-MRI, where the reconstruction with a pre-trained (global) network was used to find neighbors. 
PSNR values for the oracle LONDN-MRI reconstructor are also shown.
The overall performances with the DIDN-based architectures are better than with the UNet-based unrolled networks.
The PSNRs for LONDN-MRI are consistently and similarly better than for the globally trained network across the different training set sizes considered, indicating potential for LONDN-MRI in improving state-of-the-art models.
Fig.~\ref{fig:denoised_imgs_zoomed3} visually compares reconstructions and reconstruction errors (in zoomed in region) for different methods. 
We can see that the LONDN reconstructors capture the original image features more sharply and accurately than the globally learned reconstruction.
\begin{table}
\centering
\resizebox{0.4\textwidth}{!}{
\begin{tabular}{ccccc}
\toprule
\multicolumn{1}{c}{Acceleration}&\multicolumn{1}{c}{Data}& \multicolumn{1}{c}{Global}& \multicolumn{1}{c}{LONDN-MRI}&
\multicolumn{1}{c}{Oracle}
\\
\multicolumn{1}{c}{}& \multicolumn{1}{c}{Size}&
\multicolumn{1}{c}{}&\multicolumn{1}{c}{(1 iteration)}&\multicolumn{1}{c}{} \\
\hline
\multirow{3}{*}{$\mathrm{4x}$}  &1000& 33.66  &\textbf{33.92}    & 33.96 \\
&2000 & 34.01  & \textbf{34.23}  &34.31  \\
 & 3000&34.15  &\textbf{34.39}    & 34.42  \\
 \hline 
\multirow{3}{*}{$\mathrm{8x}$} & 1000& 31.02  & \textbf{31.33}  &31.37 \\
 & 2000&  31.34 &  \textbf{31.64}  &31.68  \\
 & 3000& 31.79  & \textbf{32.08}  &32.12  \\
\bottomrule \vspace{0.01in}
\end{tabular}}
\caption{Average reconstruction PSNR values (in dB) on the testing set at 4x and 8x undersampling for various training set sizes. MoDL reconstructor with DIDN denoiser is used. }
\label{table:PSNR_comparison_DIDN}
\end{table}

\vspace{+0.01 in}

\begin{figure*}[!t]
\centering
\begin{tabular}[b]{cccccc}
    \textbf{Ground Truth}
    &\textbf{Initial}
    & \textbf{Global} 
    & \textbf{LONDN-MRI}
    & \textbf{LONDN-MRI} 
    & \textbf{Oracle} \\
     \textbf{}
     &\textbf{}
    & \textbf{} 
    & \textbf{(1 iteration)}
    & \textbf{(2 iterations)} 
    & \textbf{}\\
    \includegraphics[width=.13\linewidth,valign=t]{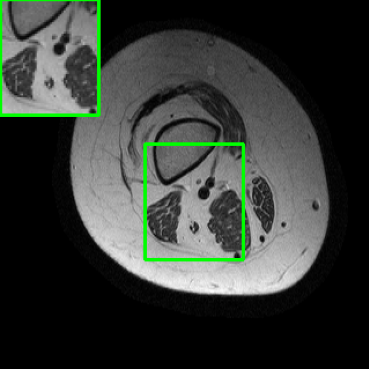} &
    \includegraphics[width=.13\linewidth,valign=t]{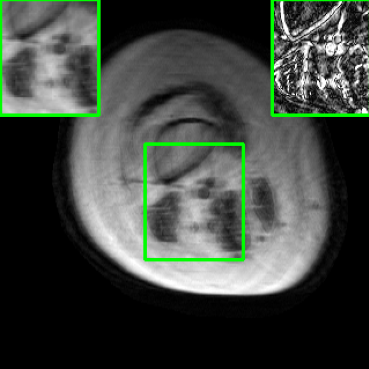} &
    \includegraphics[width=.13\linewidth,valign=t]{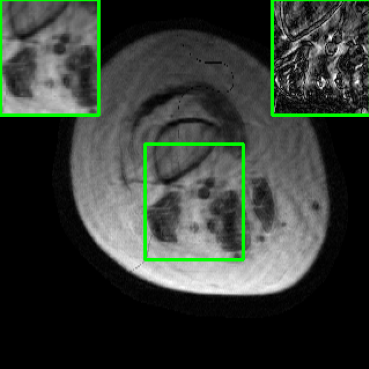} &
    \includegraphics[width=.13\linewidth,valign=t]{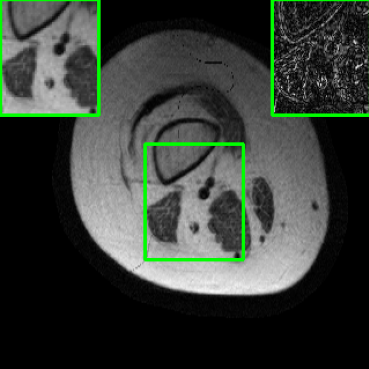} &
    \includegraphics[width=.13\linewidth,valign=t]{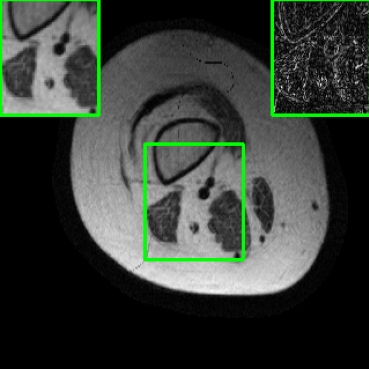} &
    \includegraphics[width=.13\linewidth,valign=t]{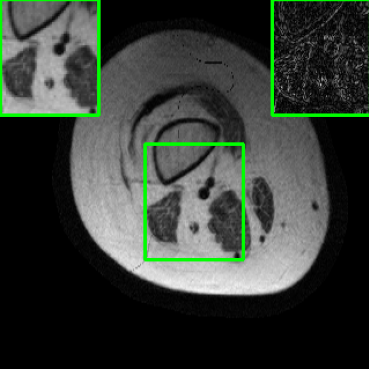} \\
    \scriptsize{PSNR = $\infty$ dB}&  \scriptsize{PSNR = 22.01 dB} & \scriptsize{PSNR = 29.02 dB} & \scriptsize{PSNR = 31.46 dB} & \scriptsize{PSNR = 31.74 dB} & \scriptsize{PSNR = 31.87 dB}
\end{tabular}%
\caption{\SR{Comparison of image reconstructions with different methods at 4x undersampling using MoDL architecture with UNet denoiser with 2000 training scans. The test slice and training data were from the Stanford FSE dataset.}}
\label{fig:denoised_imgs_zoomed_FSE}
\end{figure*}

\begin{figure*}[!t]
\centering
\begin{tabular}[b]{cccccc}
       \textbf{Ground Truth}
    &\textbf{Initial}
    & \textbf{Global} 
    & \textbf{LONDN-MRI}
    & \textbf{LONDN-MRI} 
    & \textbf{Oracle} \\
     \textbf{}
     &\textbf{}
    & \textbf{} 
    & \textbf{(1 iteration)}
    & \textbf{(2 iterations)} 
    & \textbf{}\\
    \includegraphics[width=.13\linewidth,valign=t]{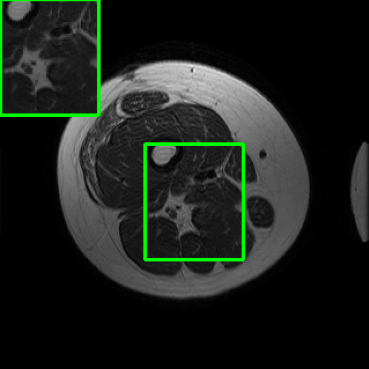} &
    \includegraphics[width=.13\linewidth,valign=t]{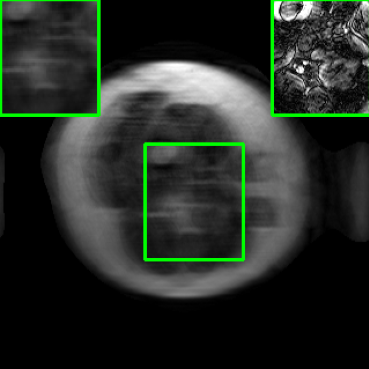} &
    \includegraphics[width=.13\linewidth,valign=t]{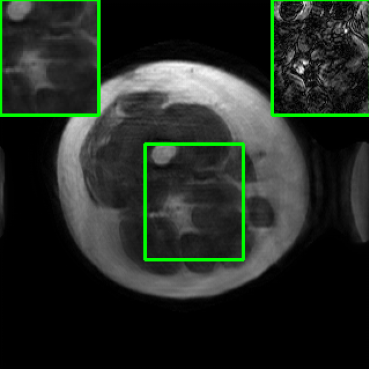} &
    \includegraphics[width=.13\linewidth,valign=t]{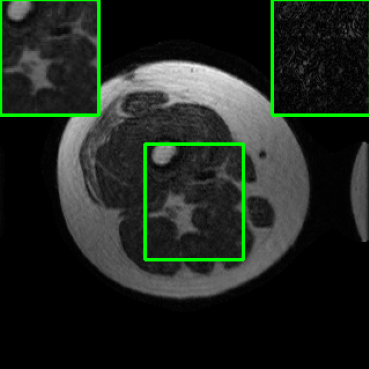} & \includegraphics[width=.13\linewidth,valign=t]{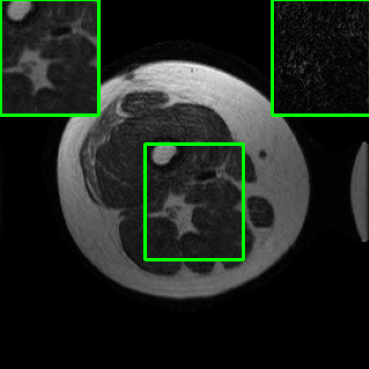} &
    \includegraphics[width=.13\linewidth,valign=t]{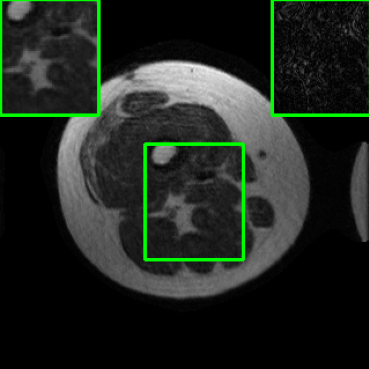}\\
    \scriptsize{PSNR = $\infty$ dB}&  \scriptsize{PSNR = 19.41 dB} & \scriptsize{PSNR = 26.52 dB} & \scriptsize{PSNR = 27.76 dB} & \scriptsize{PSNR = 27.85 dB} & \scriptsize{PSNR = 27.92 dB}
\end{tabular}%
\caption{Same comparisons/setup as Fig.~\ref{fig:denoised_imgs_zoomed_FSE}, but at 8x undersampling.}
\label{fig:denoised_imgs_zoomed_FSE_2}
\end{figure*}

\noindent \textbf{Performance on the Stanford FSE Dataset:}\label{stanfordfseresults}
\SJ{ We also performed image reconstructions with the Stanford multi-coil FSE dataset, which is a smaller dataset. 
We used same settings for the networks and training as in Section~\ref{setting}.
Table~\ref{table:PSNR_comparison_FSE} shows that LONDN-MRI significantly outperforms the globally learned MoDL network at both 4x and 8x acceleration. This indicates benefits for the proposed framework for smaller, more diverse datasets. Figs.~\ref{fig:denoised_imgs_zoomed_FSE} and~\ref{fig:denoised_imgs_zoomed_FSE_2} display visual comparisons that show the LONDN-MRI scheme recovering sharper features than the globally learned network.}

\begin{table}[htp!]
\centering
\addtolength{\tabcolsep}{-2.1pt}
\begin{tabular}{ccccc}
\toprule
\multicolumn{1}{c}{Acceleration}& \multicolumn{1}{c}{Global}& \multicolumn{1}{c}{LONDN-MRI}&
\multicolumn{1}{c}{LONDN-MRI} & 
\multicolumn{1}{c}{Oracle} \\
\multicolumn{1}{c}{}& \multicolumn{1}{p{1.5cm}}{} & \multicolumn{1}{p{1.5cm}}{(1 iteration)} &
\multicolumn{1}{p{1.5cm}}{(2 iterations)} & 
\multicolumn{1}{c}{} \\
\hline
$\mathrm{4x}$ & 29.45  & 31.49 & \textbf{31.56}   &31.67\\
$\mathrm{8x}$ & 27.25  &29.35  & \textbf{29.43}  & 29.60 \\ 
\bottomrule \vspace{0.01in}
\end{tabular}
\vspace{-0.05 in}
\caption{\SJ{Average reconstruction PSNR values (in dB) for the Stanford FSE test set at 4x and 8x  undersampling. The LONDN-MRI results are compared to a model globally trained on the FSE dataset.}}
\label{table:PSNR_comparison_FSE}
\end{table}

\subsection{Behavior of LONDN-MRI} \label{behavior}
\SR{Here, we explore the intricacies and workings of LONDN-MRI in more detail.}

\noindent \textbf{Performance with Different Distance Metrics:} \SR{To determine a suitable distance metric for our method, we analyzed a few popular distance metrics. This study focused on evaluating their effectiveness in selecting the appropriate matching dataset for training in the context of LONDN-MRI (oracle scheme). We tested the performance of MoDL with UNet denoiser using L1 and L2 distance metrics as well as normalized cross-correlation (NCC), to find the matched training set from among $3000$ images, which were all normalized. From the results in Table~\ref{tab:distancemetrics}, we see that the different distance functions offer only slight differences in reconstruction performance, with NCC offering the best results with respect to all reconstruction metrics. }

\begin{table}[ht]
    \centering
 
    \begin{tabular}{|c|c|c|c|c|}
    \hline
       Acceleration& Reconstruction Metric & L1 & L2 & NCC \\\hline
       \multirow{3}{*}{4x}&  SSIM &0.85  & 0.849 &  0.852\\\cline{2-5}
       &PSNR (dB) & 33.49 &33.44  &33.54 \\\cline{2-5}
      & HFEN &0.552 & 0.56 &0.542\\\hline
      \multirow{3}{*}{8x}& SSIM & 0.803 & 0.802 & 0.804\\\cline{2-5}
      & PSNR (dB) & 30.79 &30.71  & 30.85\\\cline{2-5}
      & HFEN & 0.664& 0.674 & 0.658\\
      \hline
    \end{tabular}
\caption{Average PSNR, SSIM, and HFEN values over 15 testing images for LONDN-MRI with neighbor search peformed using L1 distance, L2 distance, and normalized cross-correlation (NCC).}
\label{tab:distancemetrics}
\end{table}

\begin{figure}[hbt!]
\vspace{-0.01in}
\centering
\setlength{\tabcolsep}{0.5cm}
\begin{tabular}{cc}
\includegraphics[width=0.6\linewidth]{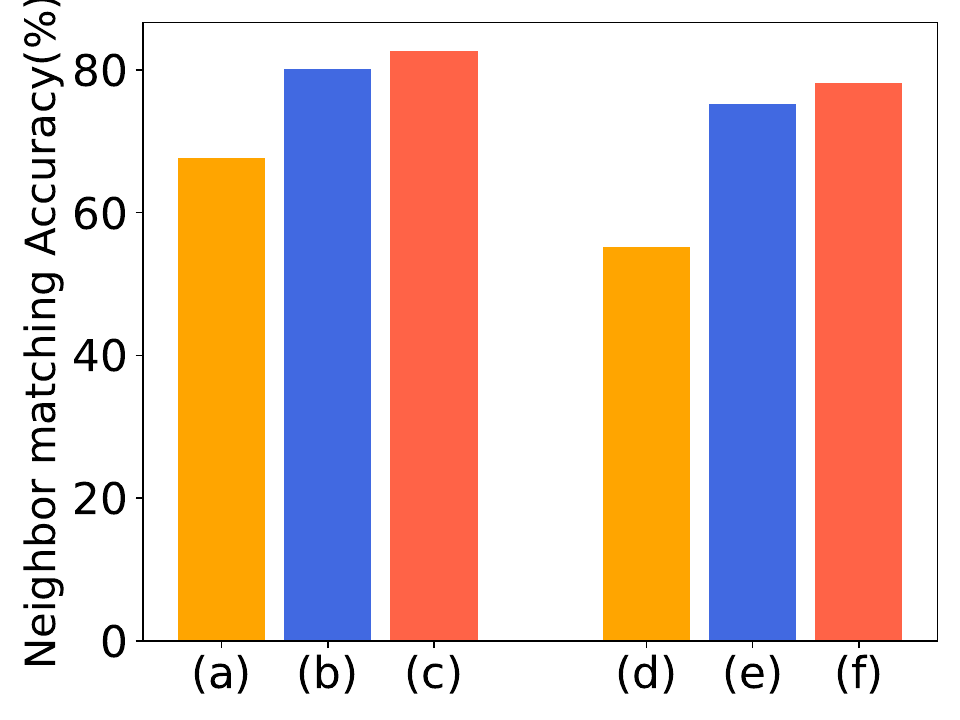}   
\end{tabular}
    \vspace{-0.05 in}
    \caption{Average accuracy (over test set) of neighbor search in LONDN-MRI (MoDL with UNet denoiser) at 4x undersampling in (a) the first iteration (neighbors found with respect to the initial input images $\x^0$) and after the (b) first and (c) second iteration. (d)-(f) are corresponding results at 8x undersampling.}
    \label{accuracy2}
\vspace{-0.05in}
\end{figure}

\noindent \textbf{Evaluating the Accuracy of Neighbor Search:} Here, we study how the neighbor search proceeds across the iterations or alternations of LONDN-MRI.
We are interested to know if our
locally learned reconstructor can improve the neighbor finding process over iterations.
We used all images from the test set.
First, we find the $k$ closest neighbors 
(in terms of Euclidean distance)
for each ground truth test image amongst the ground truth training images. The set $C_{r}^*$ contains the indices of these \emph{oracle} neighbors for a test image indexed $r$. The set $\hat{C}_{r}$ contains the indices of closest neighbors from a certain iteration of LONDN-MRI. The neighbor matching accuracy (NMA) metric below computes the average (over the test set indices $\mathcal{T}$) percentage match between the two sets:
\begin{equation}
 \text{NMA} :=   \frac{100}{|\mathcal{T}|}\sum_{r \in \mathcal{T}} \frac{|\hat{C}_{r} \cap C_{r}^*|} {k},
\end{equation}
The accuracy of the neighbor search at both 4x and 8x undersampling is shown in Fig.~\ref{accuracy2}.
The accuracy of the initial search (based on $\x^0$) and after $1$ or $2$ iterations of LONDN-MRI are shown.
We find nearest neighbors for the initial highly aliased $\x^0$ with respect to the corresponding aliased images in the training set (based on the same k-space undersampling mask as at testing time), rather than based on the ground truth training images, because the latter resulted in lower neighbor search accuracy for $\x^0$.
It is clear from Fig.~\ref{accuracy2} that
the accuracy improves quickly and tapers off in few iterations.

\begin{figure}[hbt!]
\vspace{-0.01in}
\centering
\setlength{\tabcolsep}{0.5cm}
\includegraphics[width=1.0\linewidth]{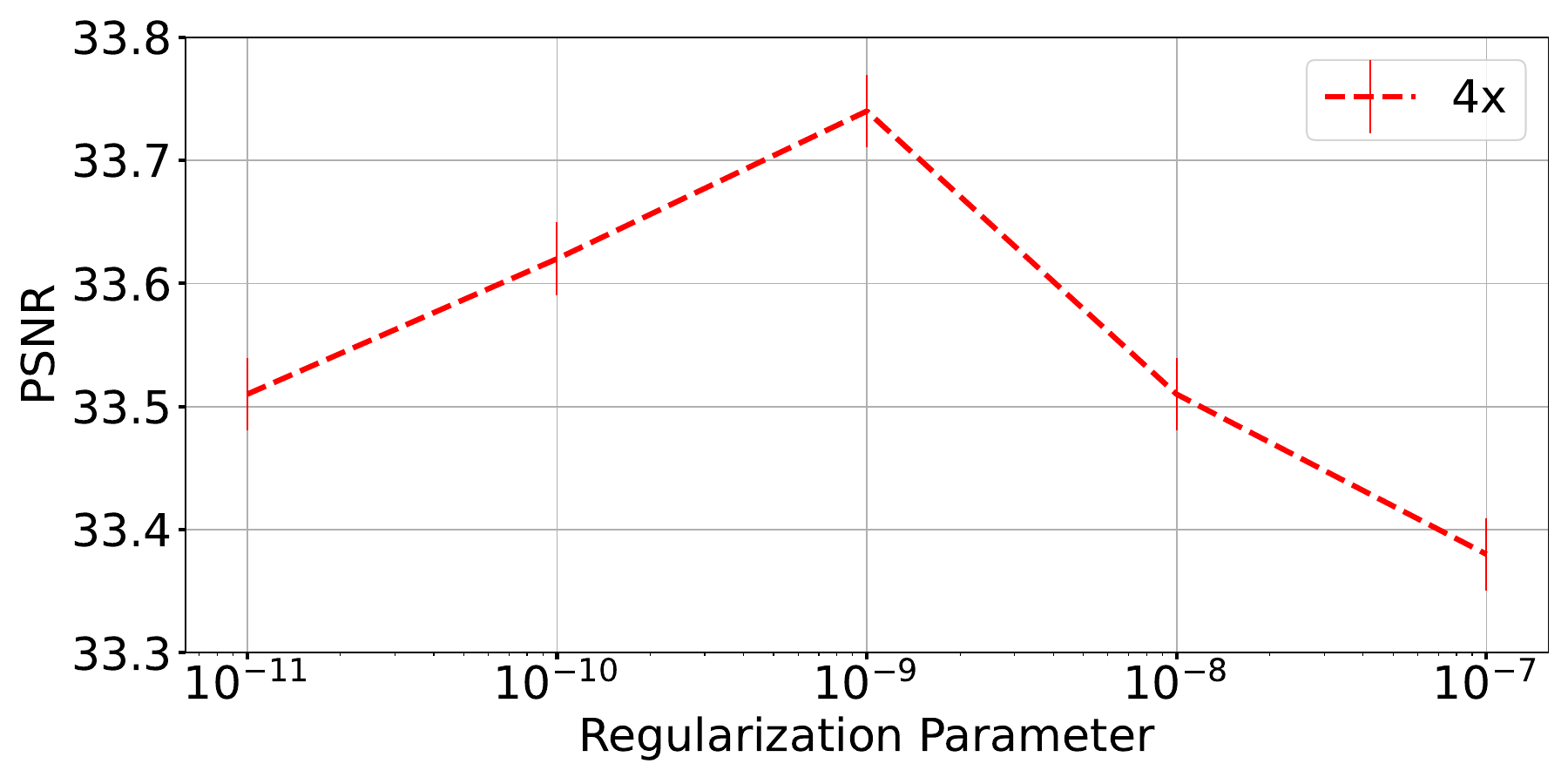}  

\caption{Average reconstruction PSNR on the test set at 4x undersampling for different regularization penalty parameters. We used $\ell_1$ norm regularization of network weights for an MoDL network with UNet denoiser.}
\label{regularization}
\end{figure}

\noindent \textbf{Effect of Weight Regularization in LONDN-MRI:} Here, we vary the strength of the regularization penalty weight in~\eqref{eq:suptrn_cost_regularization} and run LONDN-MRI over the test set at 4x k-space undersampling. Fig.~\ref{regularization} plots the average PSNR as a function of the penalty weight for the MoDL network with UNet denoiser. 
The normalized cross-correlation distance was used during neighbor search, with other parameters as before.
The result shows slight benefits for choosing the regularization weight carefully.

\begin{figure}[hbt!]
\centering
\setlength{\tabcolsep}{0.6cm}
\includegraphics[width=1.0\linewidth]{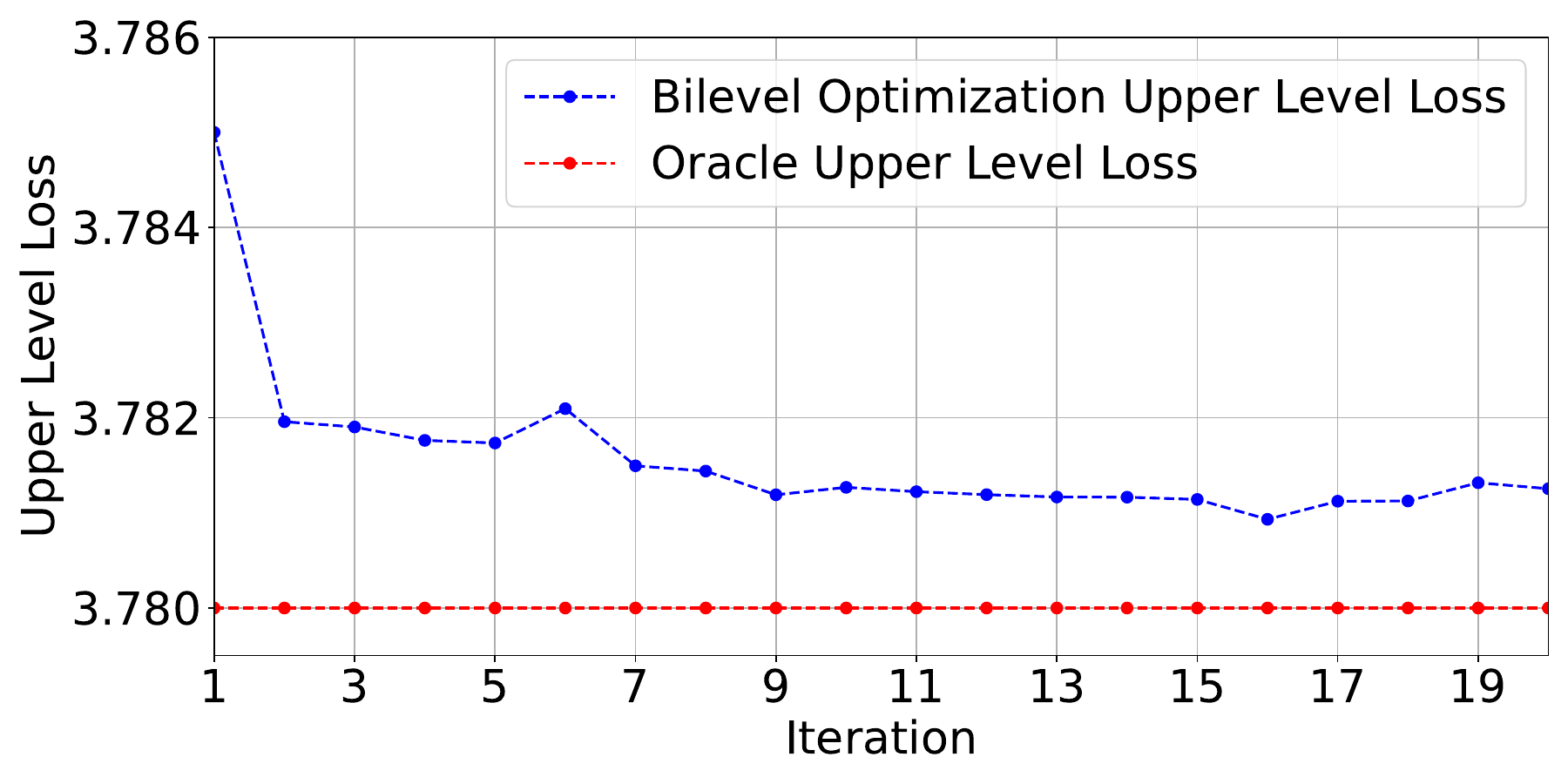}  
\caption{Upper-level loss in the bilevel optimization formulation~\eqref{eq:bilevel_problem} plotted over the iterations (after network update step) of the LONDN-MRI scheme at 4x undersampling. We used MoDL with a UNet denoiser and $k=30$ \revise{for} neighbor search. In addition, the red line shows an oracle upper-level loss computed using the ground truth test image and its $k$ nearest neighbors.}
\label{bilevel}
\end{figure}

\noindent \textbf{Convergence of Loss in Bilevel Optimization:} Next, we study the behavior of the alternating LONDN-MRI algorithm as a heuristic for the bilevel optimization formulation in~\eqref{eq:bilevel_problem}.
Here, we used an MoDL network with the UNet denoiser and $k=30$ training pairs were chosen (from 3000 cases) in the local dataset in each iteration of LONDN-MRI.
The UNet weights were randomly initialized to begin with, and the neighbor search in the first iteration of LONDN-MRI was performed using $\x^0$ and correspondingly generated aliased training images.
Fig.~\ref{bilevel} plots the upper-level loss in~\eqref{eq:bilevel_problem} (in a root mean squared error form) after each iteration of LONDN-MRI for a test image.
Here, we ran many iterations to verify convergence. We observe that the loss changes very little after a few iterations and stabilizes. This matches with the behavior of the neighbor search accuracy bar plots. The result indicates that the proposed alternating scheme could be a reasonable heuristic for reducing the loss in the challenging problem~\eqref{eq:bilevel_problem}. 
Finally, we compare the loss values in Fig.~\ref{bilevel} with an oracle loss, where the upper-level loss in~\eqref{eq:bilevel_problem} is computed using the ground truth test image and its $k$ nearest neighbors. The loss values in LONDN-MRI converge very close to the oracle loss, indicating potential for our scheme.

\begin{figure}[htb!]
\centering
\setlength{\tabcolsep}{0.02cm}
\begin{tabular}{cc}
\includegraphics[width=0.48\linewidth]{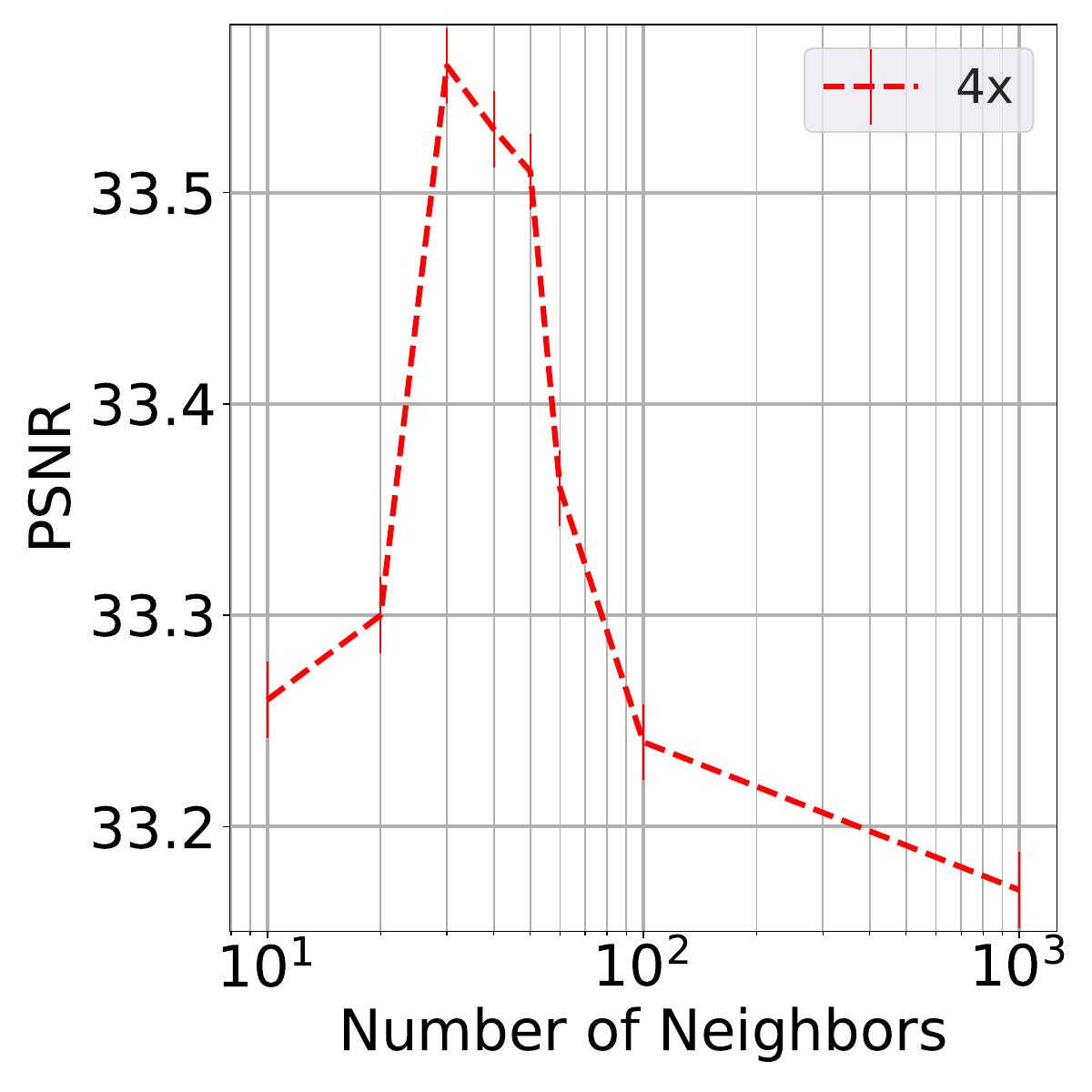}&
\includegraphics[width=0.48\linewidth]{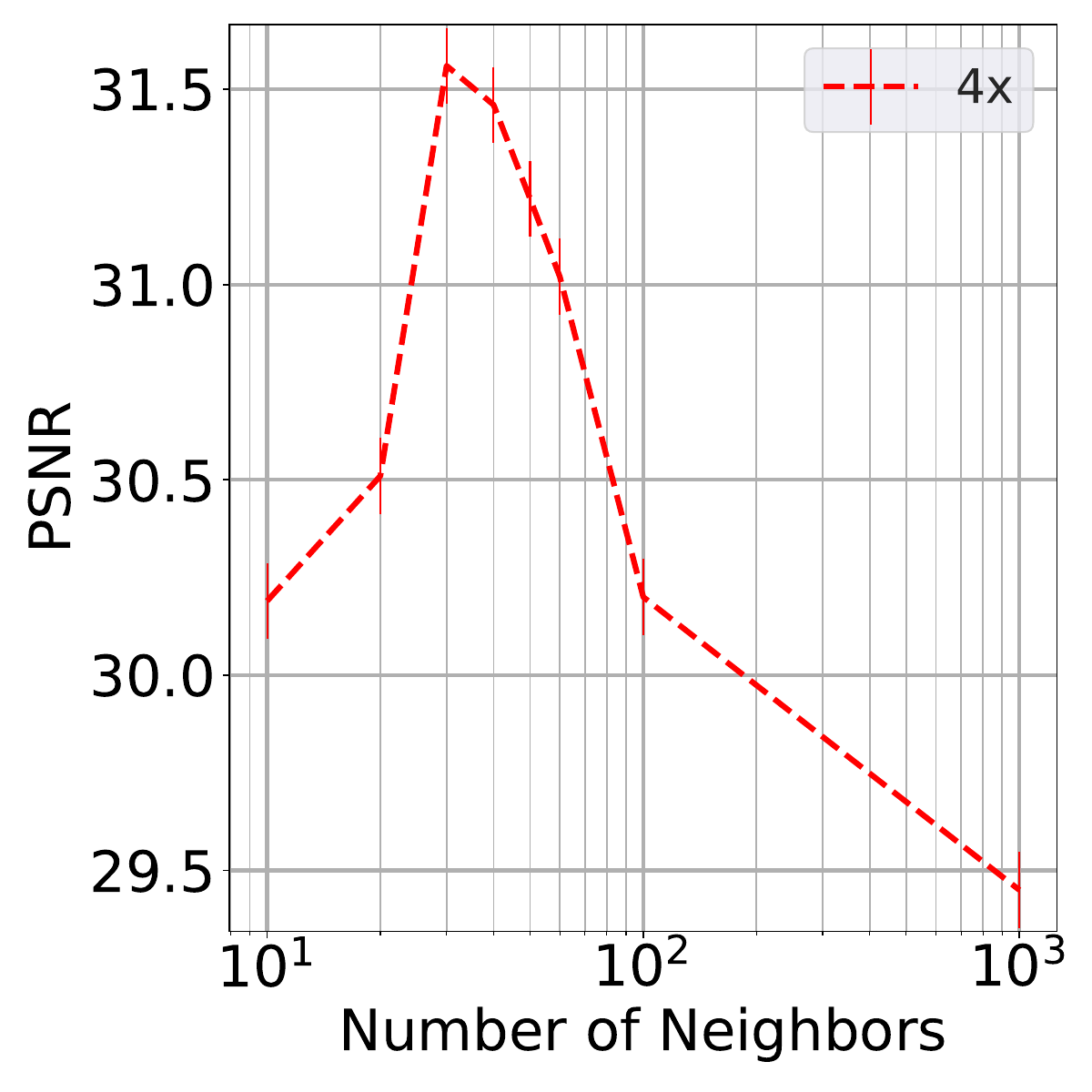}\\
(a) Fast MRI case& (b) FSE MRI case
\end{tabular}
\caption{Average reconstruction PSNR on the fastMRI and FSE MRI test set at 4x undersampling for different numbers of nearest neighbors.}
\label{neghbior reguliarzion}
\vspace*{-0.02 in}
\end{figure}

\noindent \textbf{Effect of Number of Nearest Neighbors on Image Quality:} Again, we investigate how the LONDN-MRI algorithm behaves when the number of nearest neighbors is varied to see how it affects the effectiveness of the reconstruction. To test our method, we selected from $10$ to $1000$ images for the closest neighbors (with NCC metric). The average test reconstruction PSNR for different cases is shown in Fig.~\ref{neghbior reguliarzion}. Too few local neighbors can make the method prone to overfitting and too many neighbors lead to a lack of scan-specificity and worse performance.
30-50 neighbors provide similar performances.

\begin{figure}[hbt!]
\centering
\setlength{\tabcolsep}{0.6cm}
\includegraphics[width=0.9\linewidth]{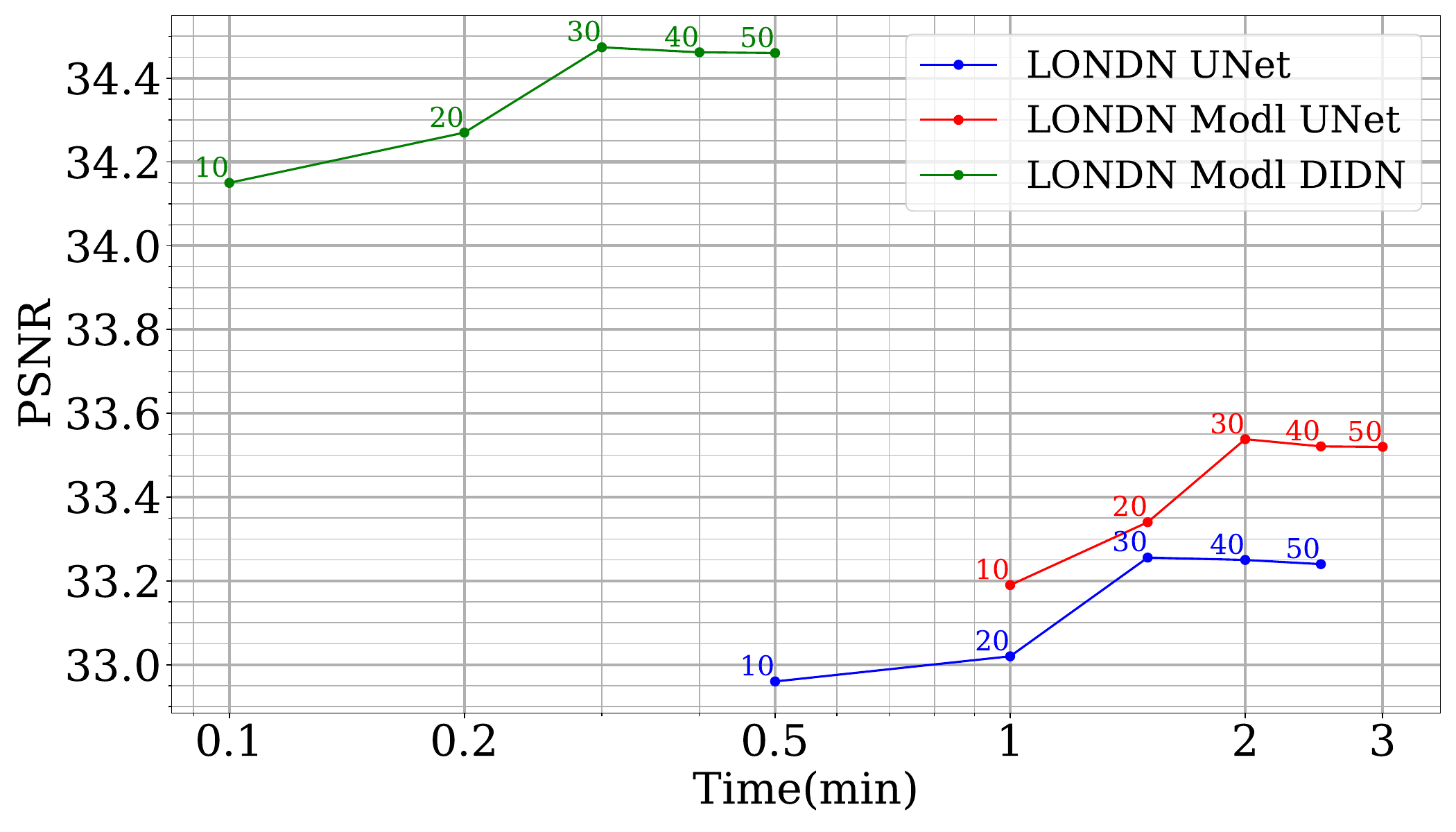}  
\caption{\SR{PSNR vs. runtime trade-offs of various LONDN-MRI models for the fastMRI knee dataset at 4x k-space undersampling. The models include MoDL networks with UNet or DIDN denoisers, as well as a standalone image-domain UNet. The performance was evaluated across different neighbor sizes, which are shown next to each data point. The processing time for these models ranged from 6 seconds to 3 minutes, depending on the neighbor size. Unrolled networks provided better image quality than the UNet denoiser.}}
\label{time_neighbor_comparsion}
\end{figure}

\noindent \textbf{Time Consumption Trade-offs:} \SR{To further understand the time efficiency of our method across different neighborhood sizes for practical applicability, we conducted comparative analyses using three models: an image-domain UNet denoiser, MODL with UNet denoiser, and the MODL with DIDN denoiser. The experiments were run on an NVIDIA GeForce RTX A5000 GPU. The PSNR vs. runtime trade-offs depicted in Figure~\ref{time_neighbor_comparsion} shed light on the time consumption for each model configuration. It is observed that some decrease in the number of neighbors leads to reduced time consumption without significantly compromising image quality. 
In addition, the results show the effectiveness of starting with a pre-trained DIDN model to improve the reconstruction, as it enhances the efficiency of the reconstruction process, reducing it to order of seconds.}

\subsection{Generalizability of LONDN-MRI}
\label{Generalizability}
\SR{Here, we present a series of studies to evaluate the generalizability of LONDN-MRI in diverse testing settings.}

\begin{figure*}
    \centering
    \begin{tabular}{cccccc}
        \textbf{Ground Truth}
    &\textbf{LORAKI}
    & \textbf{Global} 
    & \textbf{LONDN-MRI}
    & \textbf{LONDN-MRI} 
    & \textbf{Oracle}
   \\
     \textbf{}
    & \textbf{} 
    &\textbf{}
    & \textbf{(1 iteration)}
    & \textbf{(2 iterations)} 
    & \textbf{}\\
        \includegraphics[width=.13\linewidth,valign=t]{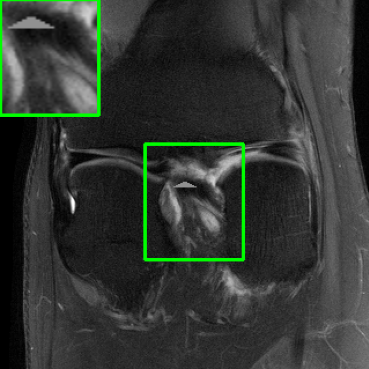} &
        \includegraphics[width=.13\linewidth,valign=t]{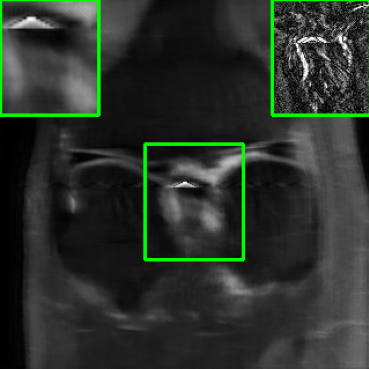} &
        \includegraphics[width=.13\linewidth,valign=t]{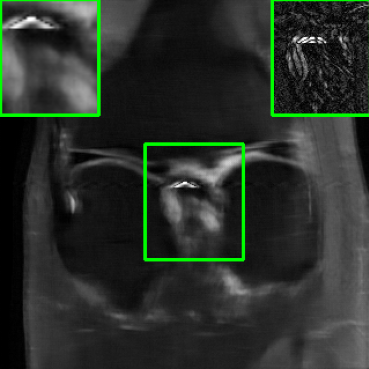} &
        \includegraphics[width=.13\linewidth,valign=t]{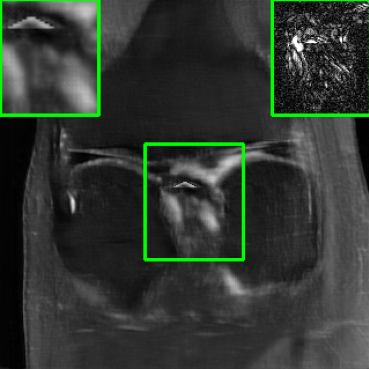}&
        \includegraphics[width=.13\linewidth,valign=t]{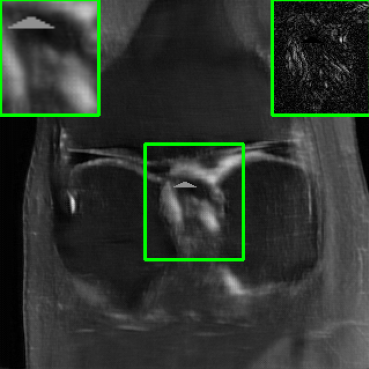} &
        \includegraphics[width=.13\linewidth,valign=t]{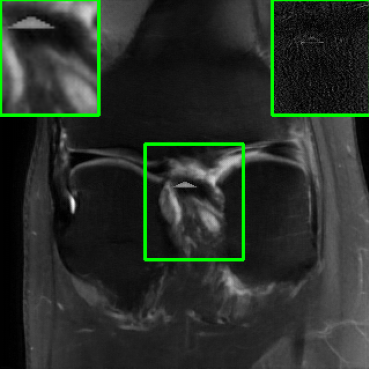} \\
        \scriptsize{PSNR = $\infty$ dB}& \scriptsize{ PSNR = 31.45 dB} &\scriptsize{ PSNR = 32.15 dB} & \scriptsize{PSNR = 32.72 dB} & \scriptsize{ PSNR = 33.15 dB}
        & \scriptsize{ PSNR = 33.26 dB}
    \end{tabular}
    \caption{\SR{Visualization of ground truth and reconstructed images using different methods at 4x k-space undersampling. The central portion (with the planted feature) and its reconstruction error map are shown in the top panels in the images.}}
    \label{fig:denoised_imgs_zoomed_inpaint}
\end{figure*}

\noindent \textbf{Performance in the Presence of Planted Features:} To assess the capability of LONDN-MRI for accurately reproducing image attributes not found in the training set (a common scenario when detecting pathologies, etc.), we embedded artificial features into a knee image from the fastMRI dataset, drawing inspiration from recent work~\cite{blipstmi2021}. We performed 4x undersampling in k-space and reconstructed with the MoDL network (with UNet denoiser) that was trained using 3000 images.
In Fig.~\ref{fig:denoised_imgs_zoomed_inpaint}, we observe that LONDN-MRI produces sharper reconstruction of image features and better PSNR compared to the globally trained network. The
details or edges of the planted features are better preserved in LONDN-MRI. Moreover, LONDN-MRI provides similar image quality with and without the planted features (Fig.~\ref{fig:denoised_imgs_zoomed2}), whereas, the globally trained network degrades significantly. This indicates the relatively improved stability and generalizability of the proposed method. 

\begin{figure*}
    \centering
    \begin{tabular}{cccccc}
    \textbf{Ground Truth}
    &\textbf{LORAKI}
    & \textbf{Global} 
    & \textbf{LONDN-MRI}
    & \textbf{LONDN-MRI} 
    & \textbf{Oracle}
   \\
     \textbf{}
    & \textbf{} 
    & \textbf{} 
    & \textbf{(1 iteration)}
    & \textbf{(2 iterations)} 
    & \textbf{}\\
        \includegraphics[width=.13\linewidth,valign=t]{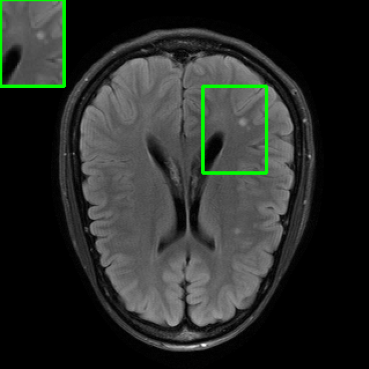}&\includegraphics[width=.13\linewidth,valign=t]{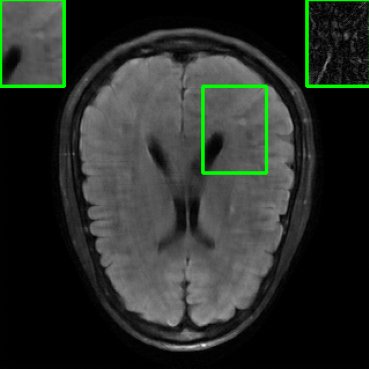}  &
        \includegraphics[width=.13\linewidth,valign=t]{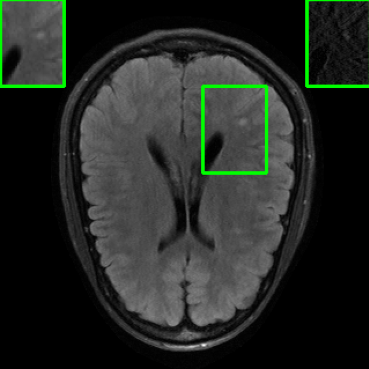} &
        \includegraphics[width=.13\linewidth,valign=t]{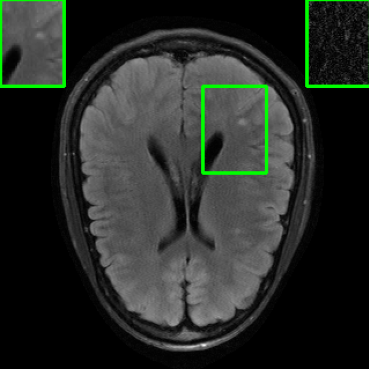} &
        \includegraphics[width=.13\linewidth,valign=t]{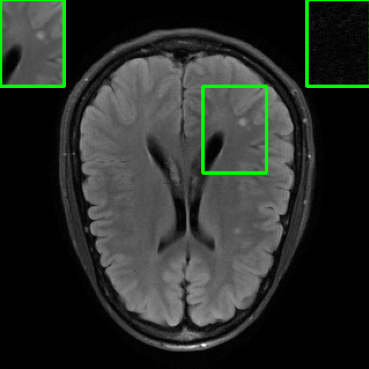} &
        \includegraphics[width=.13\linewidth,valign=t]{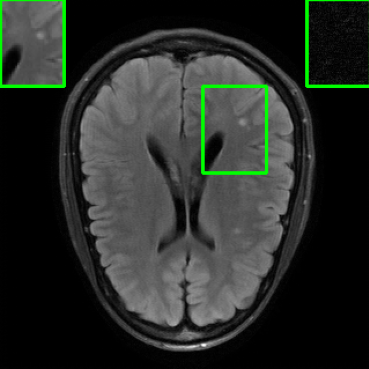} \\
        \scriptsize{ PSNR = $\infty$ dB} & \scriptsize{ PSNR = 33.17 dB} &\scriptsize{ PSNR = 35.10 dB} & \scriptsize{ PSNR = 35.67 dB}& \scriptsize{ PSNR = 35.74 dB}
        & \scriptsize{ PSNR = 35.87 dB}
    \end{tabular}
    \caption{\SR{Visualization of ground truth and reconstructed images using different methods at 4x k-space undersampling for an annotated image from the fastMRI+ dataset, where the interest area is a nonspecific white matter lesion (in green box).}}
    \label{fig:denoised_imgs_zoomed_lesion}
\end{figure*}

\noindent \textbf{Performance on Data with Lesions:}
\SR{While the previous experiment allowed comparing reconstruction quality with or without planted features, here we test our method on MRI scans with lesions, which are often regions of abnormal or diseased tissue. We utilize the annotated fastMRI+ data\footnote{{\url{https://github.com/microsoft/fastmri-plus/tree/main}}} to evaluate our method's image reconstruction capabilities, and compare its outcomes with established baselines.
For the training phase, the non-lesion dataset was employed for the global training approach with $3000$ images whereas LONDN-MRI used $30$ adaptively selected images for training (searched from $3000$ images). In contrast, during the testing phase, we used $20$ scans with lesions. The results, as displayed in Table~\ref{table:PSNR_comparison_lesion}, indicate that our method achieves substantially higher PSNR values in comparison to the globally trained baseline as well as the LORAKI method. 
Furthermore, visualizations in Figure~\ref{fig:denoised_imgs_zoomed_lesion} clearly demonstrate the superiority of our method, particularly in the nonspecific white matter lesion areas. Thus, both in terms of visual assessment and PSNR values, our approach outperforms the existing baselines and aligns more closely with the ground truth.}

\begin{table}[htp!]
\centering
\addtolength{\tabcolsep}{-2.1pt}
\resizebox{0.5\textwidth}{!}{\begin{tabular}{cccccc}
\toprule
\multicolumn{1}{c}{Acceleration}& \multicolumn{1}{c}{Global}& \multicolumn{1}{c}{LONDN-MRI}&
\multicolumn{1}{c}{LONDN-MRI} & 
\multicolumn{1}{c}{Oracle} &
\multicolumn{1}{c}{LORAKI}\\
\multicolumn{1}{c}{}& \multicolumn{1}{p{1.5cm}}{} & \multicolumn{1}{p{1.5cm}}{(1 iteration)} &
\multicolumn{1}{p{1.5cm}}{(2 iterations)} & 
\multicolumn{1}{c}{}&
\multicolumn{1}{c}{}\\
\hline
$\mathrm{4x}$ & 34.37  & 34.89 & \textbf{35.1}   &35.21&32.89\\
$\mathrm{8x}$ & 32.05  &32.65  & \textbf{32.72}  & 32.77&30.89 \\ 
\bottomrule \vspace{0.01in}
\end{tabular}}
\vspace{-0.05 in}
\caption{\SR{Average reconstruction PSNR values (in dB) for the lesion fastMRI+ test set at 4x and 8x  k-space undersampling. LONDN-MRI and the global model were trained on the non-lesion dataset.}}
\label{table:PSNR_comparison_lesion}
\end{table}

\vspace{+ 0.05 in}

\begin{figure*}
\centering
\begin{tabular}[b]{ccccc}
    \textbf{Original}& 
    \textbf{1st Nearest Neighbor}&
    \textbf{3rd Nearest Neighbor}&
    \textbf{10th Nearest Neighbor}&\\  
    \includegraphics[width=.11\linewidth,valign=t]{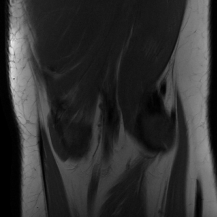}&
    \includegraphics[width=.11\linewidth,valign=t]{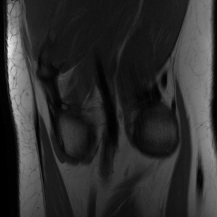}&
   \includegraphics[width=.11\linewidth,valign=t]{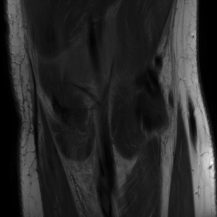} &
   \includegraphics[width=.11\linewidth,valign=t]
   {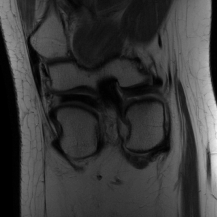}&\\
    \textbf{20th Nearest Neighbor}& 
    \textbf{30th Nearest Neighbor}&
    \textbf{40th Nearest Neighbor}&
    \textbf{50th Nearest Neighbor}&\\  
    \includegraphics[width=.11\linewidth,valign=t]{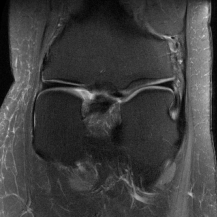}&
    \includegraphics[width=.11\linewidth,valign=t]{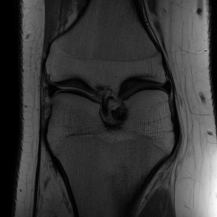}&
    \includegraphics[width=.11\linewidth,valign=t]{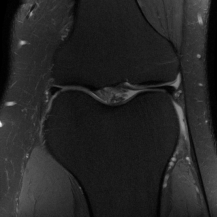}&
   \includegraphics[width=.11\linewidth,valign=t]{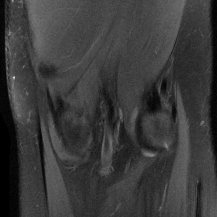}     
\end{tabular}
\caption{\SL{An image is shown along with different nearest neighbors from the fastMRI dataset.}}
\label{fig:Imgs_neighbor}
\end{figure*}

\noindent \textbf{Performance without Well-Matched Neighbors:} \SL{Another natural question is how sensitive is the proposed method to using a `well matched' (to the test scan)  subset of images in the global training set. 
One might consider this restrictive.
To better evaluate the working of LONDN-MRI, we switched its training with the UNet denoiser from using the $30$ closest neighbors to using the 31st to the 60th closest (or less similar) neighbors.
Fig.~\ref{fig:Imgs_neighbor} shows an example with the different near-neighbors that are chosen from the $3000$ image global training set, ranked based on NCC distance. While the nearest neighbors look quite similar to the test image, the farther ones could be relatively dissimilar in practice.
In this case, LONDN-MRI (with 1 iteration) using the 31st to the 60th closest neighbors still reconstructs the test scans well with an average PSNR of $33.34$ dB (at 4x k-space undersampling and 15 test images), which is only slightly worse than when using the $30$ closest neighbors ($33.46$ dB).
This indicates the proposed approach may not be very sensitive to availability of highly visually matched training data.
Indeed, the Stanford FSE data has more variability than fastMRI and our approach performs well on that dataset.}

\noindent \textbf{Evaluating Generalization with Limited Training Sets:}
\SR{To facilitate a fairer comparison with 
scan-adaptive methods such as DIP, LORAKI and RAKI, we conduct experiments utilizing much smaller subsets of the original fastMRI knee dataset, from which the neighbors in LONDN-MRI are selected.
We randomly selected $5$ to $100$ slices for the overall training set in LONDN-MRI. These were chosen from a small random set of volumes/patients. The goal is to emulate comparisons with DIP, LORAKI, and RAKI when LONDN-MRI operates in a very limited dataset regime.}
For each overall training set size, we selected the top $k$ similar neighbors at testing time, where $k$ is adjusted based on the dataset size. For example, for a dataset with $5$ slices, we selected the top $3$ similar scans at test time, and for a dataset with $100$ samples, we selected the top $10$ neighbors in the search.

\begin{figure}[hbt!]
\centering
\setlength{\tabcolsep}{0.6cm}
\includegraphics[width=0.9\linewidth]{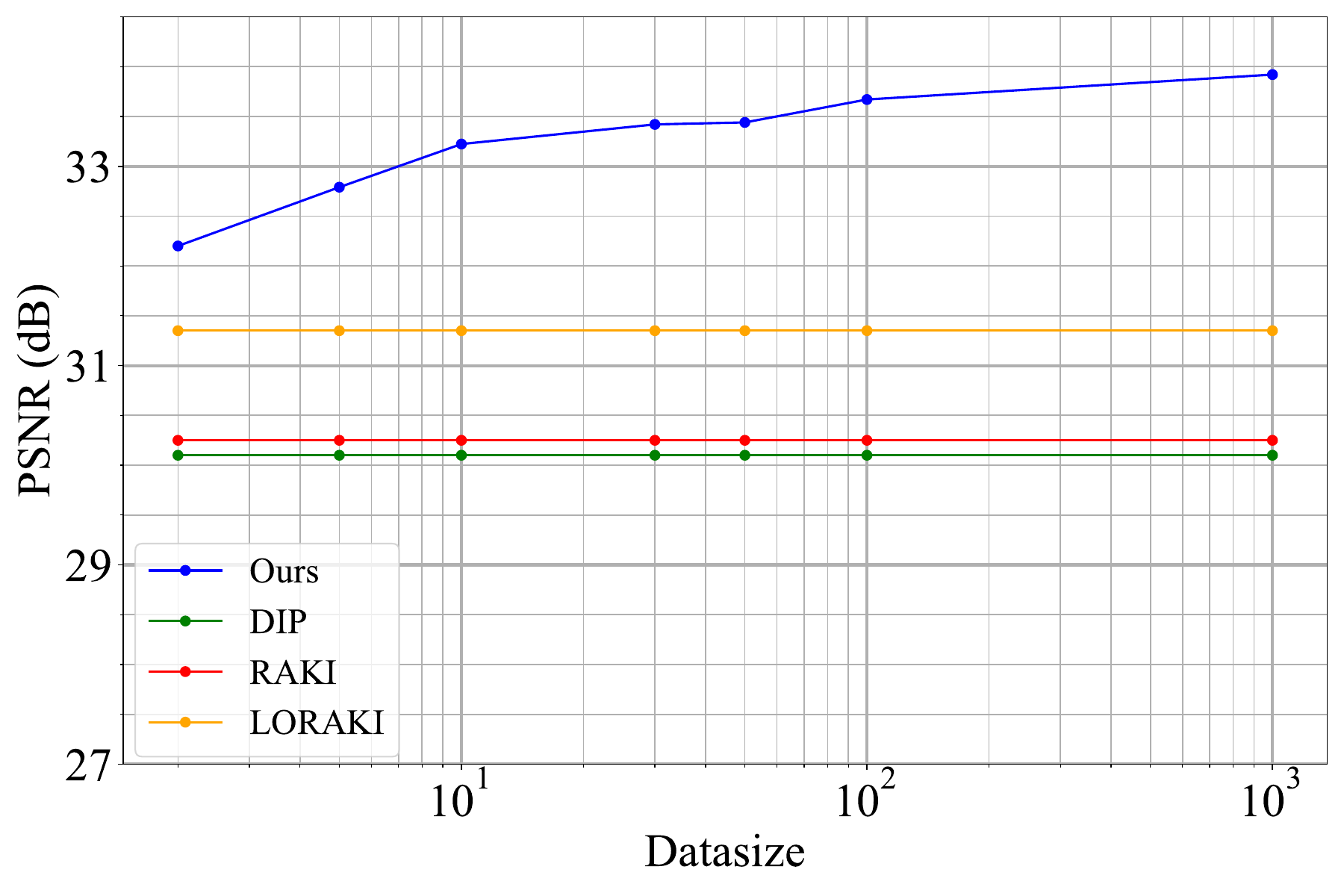}  
\caption{\SR{Average PSNR on test set (from fastMRI) for LONDN-MRI (MoDL network with UNet denoiser) at 4x k-space undersampling for various  dataset sizes. Subsets of the dataset are chosen as neighbors in LONDN-MRI at test time. The average PSNR values with DIP, LORAKI, and RAKI, which require no training data are shown as horizontal lines.}}
\label{neighbor_comparsion}
\end{figure}

The average reconstruction PSNR for the testing scans,
plotted in Fig.\ref{neighbor_comparsion}, reveals that although there is some decline in performance with decreasing dataset size, the results still surpass those achieved by DIP, RAKI and LORAKI, indicating potential for LONDN-MRI with very limited training sets.
While DIP, RAKI and LORAKI adapt purely to the individual test scans without supervision,
the LONDN-MRI approach wouldn’t make sense in the 0-paired data regime. In future work, we plan to study hybrid methods leveraging both LONDN-MRI and DIP, i.e., adapting the network based on both similar paired data and the current test scan’s measurements (as in DIP).

\begin{figure*}
    \centering
    \begin{tabular}{cccccc}
    \textbf{Ground Truth}
    &\textbf{LORAKI}
    & \textbf{Global} 
    & \textbf{LONDN-MRI}
    & \textbf{LONDN-MRI} 
    & \textbf{Oracle}
   \\
     \textbf{}
    & \textbf{} 
    & \textbf{} 
    & \textbf{(1 iteration)}
    & \textbf{(2 iterations)} 
    & \textbf{}\\
        
        \includegraphics[width=.13\linewidth,valign=t]{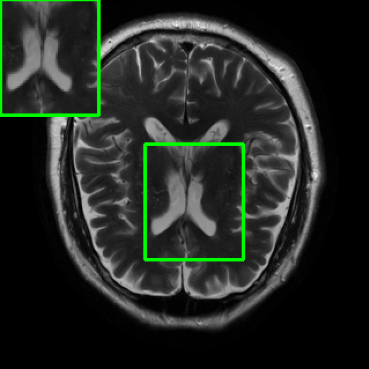} &        \includegraphics[width=.13\linewidth,valign=t]{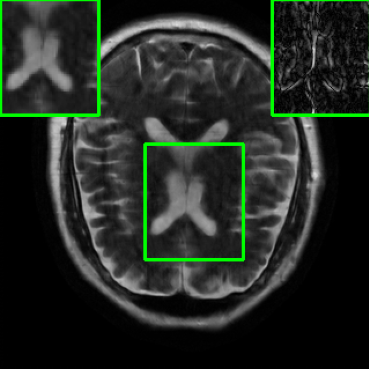}&\includegraphics[width=.13\linewidth,valign=t]{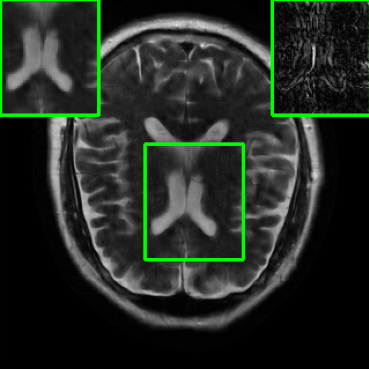} &
        \includegraphics[width=.13\linewidth,valign=t]{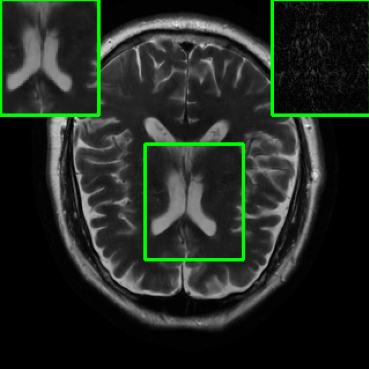} &
        \includegraphics[width=.13\linewidth,valign=t]{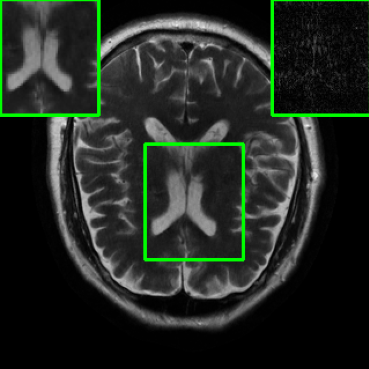}&
        \includegraphics[width=.13\linewidth,valign=t]{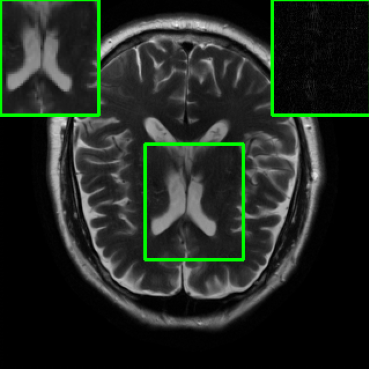}  \\
        \scriptsize{ PSNR = $\infty$ dB} & \scriptsize{ PSNR = 33.21 dB} & \scriptsize{ PSNR = 36.12 dB} & \scriptsize{ PSNR = 36.31 dB}&
        \scriptsize{ PSNR = 36.54 dB}&
        \scriptsize{ PSNR = 36.77}
        
    \end{tabular}
    \caption{\SR{Visualization of ground-truth and reconstructed images using different methods at 4x k-space undersampling for a T2 contrast MRI scan (with training on T1 contrast scans). A region of interest (in green box) and its error map are also shown.}}
    \label{fig:denoised_imgs_zoomed_T2}
\end{figure*}
\vspace{+0.05in}

\noindent \textbf{Effect of Varying Scan Settings at Test Time:} Since the reconstruction network in LONDN-MRI is trained for each scan, we would like to understand better the benefits this provides in terms of letting the network adapt to distinct scan settings. So we chose the MoDL reconstructor with UNet denoiser (with same hyperparameters for training as before) and trained it on the 3000 image set in two ways: with a fixed sampling mask across the images (the mask was padded with zeros to account for slight variations in matrix sizes), and with a different random sampling mask for each image. The \revise{first} setting was used in previous subsections. For LONDN-MRI, here, we used a different random sampling mask for each test scan, but the network was adapted locally with the same mask used across each (small) local training set.
Table~\ref{table:PSNR_comparison_mask} shows the average PSNR values on the test set with these different strategies as well as with the oracle LONDN-MRI scheme.
It is clear that the globally learned model with a fixed sampling mask struggles to generalize to the different scan settings at test time. But training the global model with random sampling masks leads to improved reconstruction PSNRs.
Importantly, the LONDN-MRI schemes that adapt the reconstruction model to the settings as well as the data for each scan provide marked improvements over both globally learned network settings.

\begin{table}[htp!]
\centering
\addtolength{\tabcolsep}{-2.1pt}
\begin{tabular}{ccccc}
\toprule
\multicolumn{1}{c}{Acceleration}& \multicolumn{1}{c}{Global Model}& \multicolumn{1}{c}{Global Model}&
\multicolumn{1}{c}{LONDN-MRI} & 
\multicolumn{1}{c}{Oracle} \\
\multicolumn{1}{c}{}& \multicolumn{1}{p{1.5cm}}{trained with a fixed mask} & \multicolumn{1}{p{1.5cm}}{trained with rand. masks} &
\multicolumn{1}{p{1.5cm}}{(2 iterations)} & 
\multicolumn{1}{c}{LONDN} \\
\hline
$\mathrm{4x}$ & 33.03  & 33.19 & \textbf{33.56}   &33.64\\
$\mathrm{8x}$ & 30.62  & 30.84 &\textbf{31.14}  & 31.22 \\
\bottomrule \vspace{0.01in}
\end{tabular}
\vspace{-0.05 in}
\caption{Average reconstruction PSNR values (in dB) on the test set at 4x and 8x undersampling. The LONDN-MRI results are compared to training a global model with a fixed sampling mask or with random masks.}
\label{table:PSNR_comparison_mask}
\end{table}

\begin{figure}[hbt!]
\centering
\setlength{\tabcolsep}{0.6cm}
\includegraphics[width=1.0\linewidth]{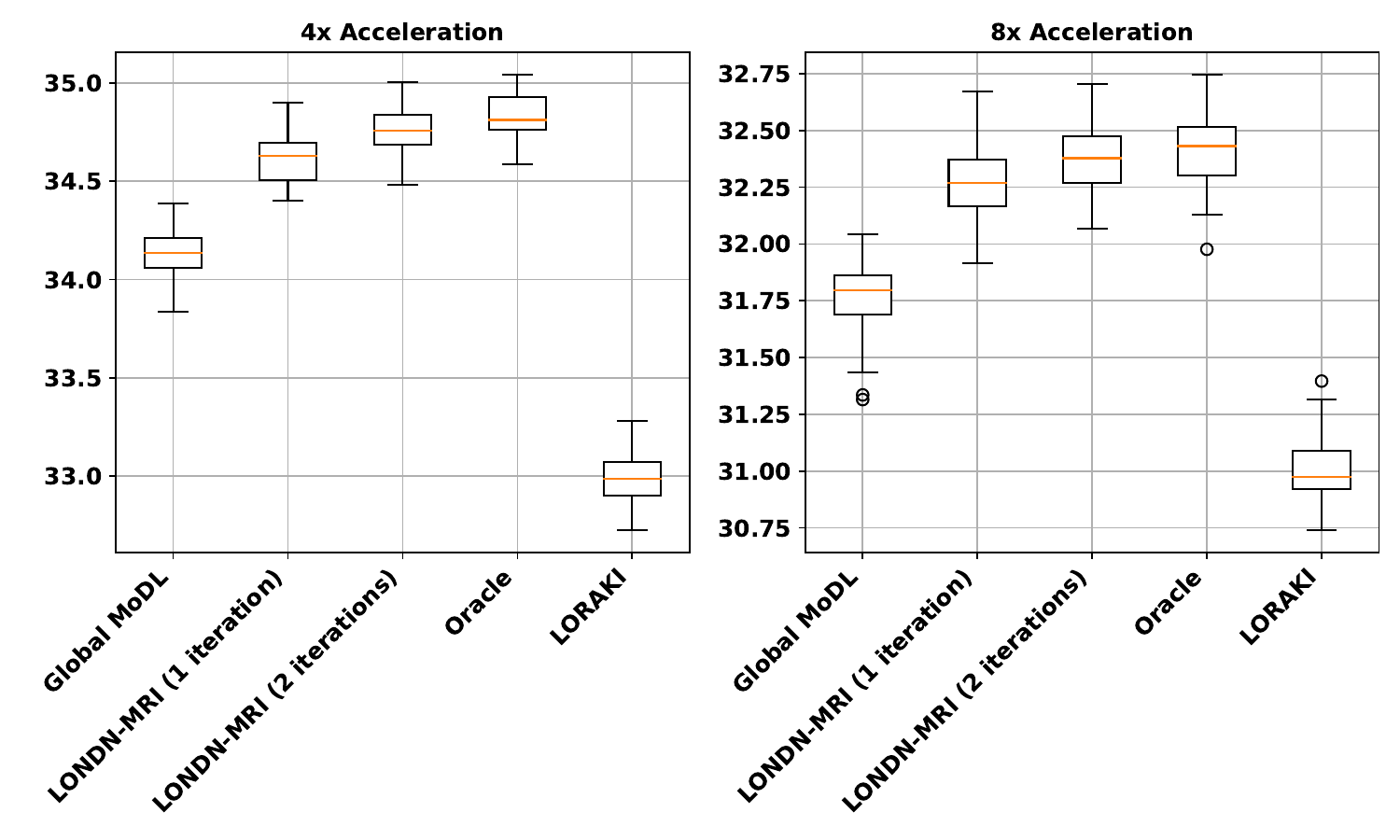}  
\caption{\SR{Box plots for average reconstruction PSNR values (in dB) for different methods for the T2 fastMRI brain test set at 4x and 8x undersampling. LONDN-MRI (trained on T1 contrast fastMRI dataset)  results are compared to a model trained globally (on $3000$ T1 contrast scans) and to LORAKI.}}
\label{boxplot_brain}
\end{figure}

\noindent \textbf{Results with Different Contrasts:} \SR{To delve deeper into clinical applicability of our method, we conducted further tests to ascertain its adaptability to different contrasts or weightings in scans. Conventional deep learning reconstruction techniques may need consistency in contrast between training and testing to achieve optimal results and could struggle with generalization across varied experimental settings. Our method, being scan-specific, could offer some flexibility because of adaptivity to features in test scans. To further study this, we conducted a test, where the global model was trained exclusively on T1 MRI data at 4x and 8x undersampling
using $3000$ training scans. Subsequent testing was done on T2 contrast MRI data with $20$ images. For LONDN-MRI, we used $30$ images for local training (searched from $3000$ images) for each test scan. The results, presented as box plots in Fig.~\ref{boxplot_brain} and visualized with one example in Fig~\ref{fig:denoised_imgs_zoomed_T2}, highlight our method's reconstruction performance in comparison to the globally trained MoDL network and the scan-specific LORAKI scheme. 
Our method exhibits notable better performance, underscoring its effectiveness in diverse imaging contexts.}   

\begin{figure}[hbt!]
\centering
\setlength{\tabcolsep}{0.6cm}
\includegraphics[width=1.0\linewidth]{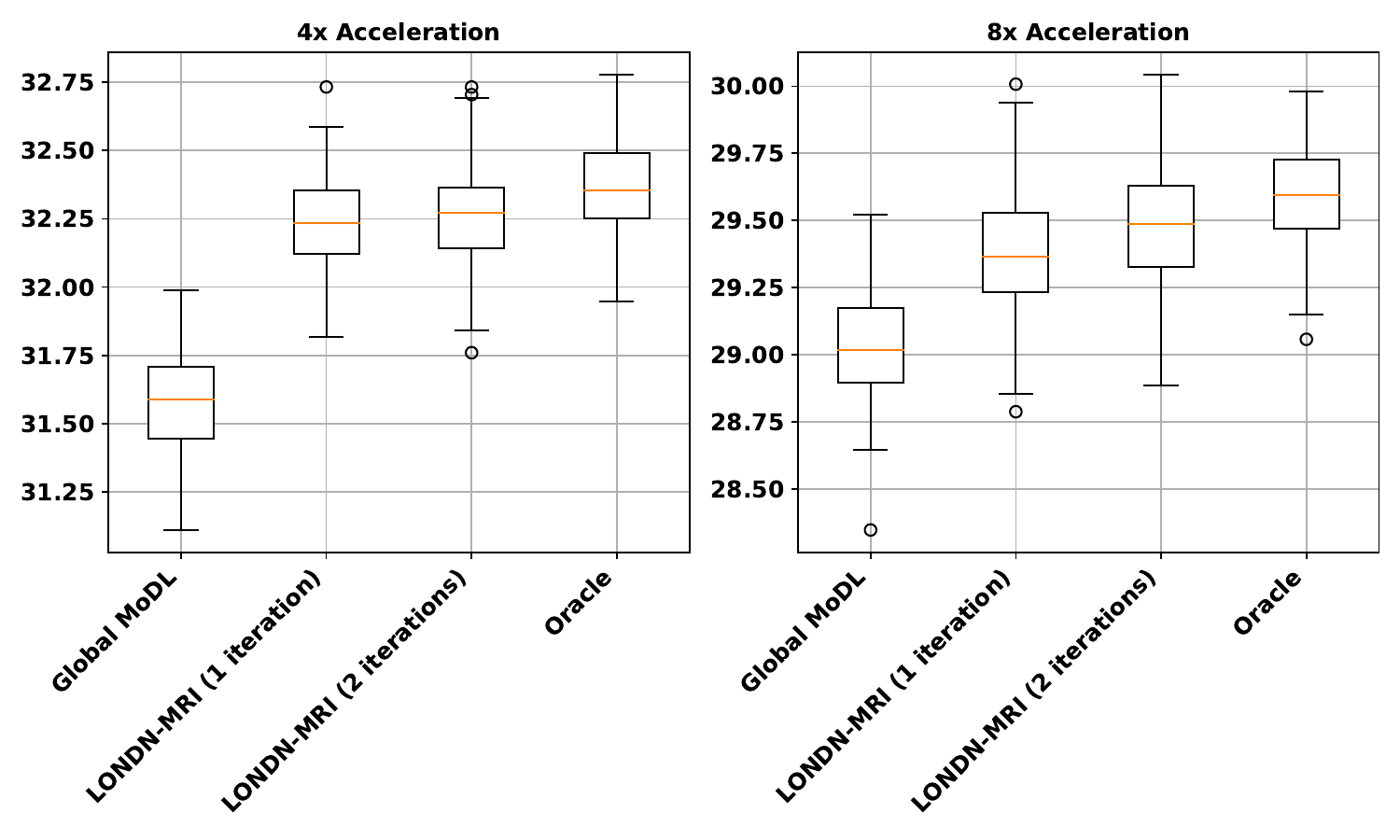}  
\caption{\SR{Box plots of average reconstruction PSNR values (in dB) for different methods on the fastMRI knee test set at 4x and 8x  undersampling. For the test dataset, we added zero-mean Gaussian noise to the measurements with standard deviation $\sigma = 0.01$ for the real and imaginary parts of the noise. All training data used did not include additional noise.}}
\label{boxplot_noise}
\end{figure}

\noindent \textbf{Performance with Different Signal-to-Noise Ratios:} \SR{To assess the performance of LONDN-MRI when the training and tested data have different signal-to-noise ratios (SNRs), we conducted tests on scans from the fastMRI knee dataset that were subjected to additive random Gaussian noise with a variance of $0.01$ for the real and imaginary parts of the noise. The globally and locally trained models at 4x and 8x undersampling used data without added noise, and the training settings were the same as before in Section~\ref{setting}. Our findings revealed a general decline in reconstruction performance across all methods, attributable to the different SNRs between training and testing. Despite this, LONDN-MRI displays better capability in handling noise perturbations, with a wider performance gap over the globally trained model. This is clear from the PSNR values depicted in the corresponding box plots in Fig.~\ref{boxplot_noise}.}

\section{Discussion} \label{section5}
We proposed a novel LONDN-MRI reconstruction technique that efficiently matches test reconstructions to a cluster of a dataset, where networks are adaptively  estimated on images most related to a current scan.
\SR{Our results on the multi-coil fastMRI brain and knee datasets, fastMRI+, and the Stanford FSE dataset showed promise for our patient-adaptive network estimation scheme.} The approach does not require pre-training and can thus readily handle changes in the training set. 
Additionally, the networks in LONDN-MRI can be randomly initialized and trained adaptively on very small datasets, and such networks outperformed models trained globally on much larger datasets (with lengthy training times).
For example, for fastMRI knee scans,
LONDN-MRI with $2$ alternations involving MoDL with a randomly initialized UNet denoiser took $5$ minutes to run on a NVIDIA GeForce RTX A5000 GPU (with batchsize of $6$ and $200$ epochs each time to update networks locally).
\SR{While LONDN-MRI 
outperformed the scan-adaptive 
methods such as DIP, RAKI, and LORAKI in image quality, the runtimes for the methods were somewhat similar.} \SR{DIP takes about $5$ minutes to reach peak performance (over iterations) with the same GPU, while RAKI and LORAKI took $3$ mins and $4$ mins, respectively.} 
\SR{LONDN-MRI requires only a few images (e.g., $30$) to train networks, with often 200-250 epochs for locally updating randomly initialized networks such as the UNet. Fewer epochs (often $10$ suffices) of update were needed with pre-trained networks such as the pre-trained DIDN, resulting in runtimes of only 18 seconds per iteration of LONDN-MRI (Fig.~\ref{time_neighbor_comparsion}). Of course, a globally trained model would run faster at inference time. For example, MoDL with pre-trained DIDN denoiser takes $8$ seconds on average to reconstruct fastMRI knee images.} 
\SL{Note that the neighbor search process in the proposed method is highly efficient. We find 20-30 images from $3000$ images to train the model in about $10$ seconds, while the overall algorithm takes minutes. The neighbor search is also highly parallelizable.}


When compared to the supervised global model, the proposed method  offers consistently improved reconstruction quality in terms of PSNR, SSIM, and HFEN metrics.
\SR{Additionally, we demonstrated that the local model adapts better to test time changes 
(such as changes to the sampling mask, scan contrast, SNR, presence of anomalies, etc.)
compared to a globally learned (and fixed) model.} 
\SR{Our approach produced marked improvements for the Stanford FSE dataset, and noticeable improvements for fastMRI/fastMRI+.} 
Additionally, our study with different distance metrics revealed they have only slight effect on reconstruction quality. 
\SR{The NCC metric provided the best reconstruction quality and was thus used in our studies.} We conjecture that a learned distance metric~\cite{sym11091066} could further enhance the performance of LONDN-MRI.

\section{Conclusions} \label{section6}

This paper examined supervised learning of deep unrolled networks at reconstruction time for MRI by exploiting training sets along with local modeling and clustering.
We showed advantages for this approach at different k-space undersampling factors over networks learned in a global manner on larger data sets. The training may be connected to a bilevel optimization problem.
We also compared different distance metrics for finding neighbors in our approach and regularization to reduce local overfitting.
\SR{We intend to expand our studies in the future by incorporating non-Cartesian undersampling patterns, such as radial and spiral patterns, as well as deploying them to 3D settings and other imaging modalities.}
Additionally, the method's generalizability will be further examined, with a particular emphasis on heterogeneous datasets. \textcolor{black}{To handle more extreme training-test data variations such as unseen anatomies, we plan to explore patch-based neighbors in local learning schemes for future work. }
We showed benefits for both randomly seeded training of simple models and for fine tuning of sophisticated pre-trained models, and
\SL{believe our methodology could be applied to a variety of deep learning-based 
tasks (even beyond image reconstruction)} effectively
to improve overall performance. 
Finally, metric learning~\cite{sym11091066} to improve local clustering and subsequent network adaptation will be an important future direction.

{\small
\bibliographystyle{IEEEbib}
\bibliography{TCI_arxiv}
}

\end{document}

%% file: macro.tex
\newcommand{\argmin}[1] {\underset{#1}{\textnormal{arg min}}}

\newcommand{\xmath}[1] {\ensuremath{#1}\xspace}
\newcommand{\blmath}[1] {\xmath{\bm{#1}}}
\newcommand{\paren}[1]{\xmath{\left( #1 \right)}}

\newcommand{\defequ}{\triangleq}
\newcommand{\revise}[1]{\textcolor{black}{#1}}

\newcommand{\D}{\xmath{\bm{D}}}

\newcommand{\z}{\xmath{\bm{z}}}

\newcommand{\x}{\xmath{\bm{x}}}

\newcommand{\y}{\xmath{\bm{y}}}

\newcommand{\Sb}{\xmath{\bm{S}}}
\newcommand{\A}{\blmath{A}}

\newcommand{\Sbs}[2]{\Sb^{#1}_{#2}}
\newcommand{\scrM}{\xmath{\bm{\mathscr{M}}}}

\newcommand{\respp}[1]{\marginpar{\textcolor{blue}{}}}
\newcommand{\resp}[1]{\marginpar{\textcolor{blue}{}}}

\makeatletter
\newcommand{\bigcomp}{%
  \DOTSB
  \mathop{\vphantom{\sum}\mathpalette\bigcomp@\relax}%
  \slimits@
}
\newcommand{\bigcomp@}[2]{%
  \begingroup\m@th
  \sbox\z@{$#1\sum$}%
  \setlength{\unitlength}{0.9\dimexpr\ht\z@+\dp\z@}%
  \vcenter{\hbox{%
    \begin{picture}(1,1)
    \bigcomp@linethickness{#1}
    \put(0.5,0.5){\circle{1}}
    \end{picture}%
  }}%
  \endgroup
}
\newcommand{\bigcomp@linethickness}[1]{%
  \linethickness{%
      \ifx#1\displaystyle 2\fontdimen8\textfont\else
      \ifx#1\textstyle 1.65\fontdimen8\textfont\else
      \ifx#1\scriptstyle 1.65\fontdimen8\scriptfont\else
      1.65\fontdimen8\scriptscriptfont\fi\fi\fi 3
  }%
}
\makeatother